\documentclass[manuscript]{aastex}

\usepackage[percent]{overpic}
\usepackage{pstricks}
\usepackage{tikz}

\newlength{\mywidth}
\newlength{\myheight}

\newcommand{\pgfsize}[2]{ 
 \pgfextractx{#1}{\pgfpointdiff{\pgfpointanchor{current bounding box}{south west}}
 {\pgfpointanchor{current bounding box}{north east}}}
 \pgfextracty{#2}{\pgfpointdiff{\pgfpointanchor{current bounding box}{south west}}
 {\pgfpointanchor{current bounding box}{north east}}}
}

\def \widthoffset {5}
\def \heightoffset {23}
\def \plotwidth {5cm}

\shorttitle{XEUV in Disks}
\shortauthors{Pascucci et al.}

\newcommand{\neii}{[Ne~{\sc ii}]}
\newcommand{\neiii}{[Ne~{\sc iii}]}
\newcommand{\oi}{[O~{\sc i}]}

\begin{document}


\title{Low EUV Luminosities Impinging on Protoplanetary Disks}


\author{I. Pascucci}
\affil{Lunar and Planetary Laboratory, The University of Arizona, Tucson, AZ 85721, USA}
\email{pascucci@lpl.arizona.edu}

\author{L. Ricci}
\affil{Department of Astronomy, California Institute of Technology, MC 249-17, Pasadena, CA 91125, USA}

\author{U. Gorti\altaffilmark{1} and D. Hollenbach}
\affil{SETI Institute, 189 Bernardo Ave., Mountain View, CA 94043, USA}
\altaffiltext{1}{NASA Ames Research Center, Moffett Field, CA 94035, USA}

\author{N. P. Hendler}
\affil{Lunar and Planetary Laboratory, The University of Arizona, Tucson, AZ 85721, USA}

\and
\author{K. J. Brooks and Y. Contreras}
\affil{Australia Telescope National Facility, PO Box 76, Epping, NSW 1710, Australia}




\begin{abstract}
The amount of high-energy stellar radiation reaching the surface of protoplanetary disks is essential to determine their chemistry and physical evolution. Here, we use millimetric and centimetric radio data to constrain the EUV luminosity impinging on 14 disks around young ($\sim$2-10\,Myr) sun-like stars. For each object we identify the long-wavelength emission in excess to the dust thermal emission, attribute that to free-free disk emission, and thereby compute an upper limit to the EUV reaching the disk. We find upper limits lower than 10$^{42}$\,photons/s for all sources without jets and lower than $5 \times 10^{40}$\,photons/s for the three older sources in our sample. 
These latter values are low for EUV-driven photoevaporation alone to clear out protoplanetary material in the timescale inferred by observations. In addition, our EUV upper limits are too low to reproduce 
the \neii{} 12.81\,\micron{} luminosities from three disks with slow \neii-detected winds. This indicates that the \neii{} line in these sources primarily traces a mostly neutral wind where Ne is ionized by 1\,keV X-ray photons, implying higher photoevaporative mass loss rates than those predicted by EUV-driven models alone. In summary, our results suggest that high-energy stellar photons other than EUV may dominate the dispersal of protoplanetary disks around sun-like stars.
\end{abstract}





\section{Introduction}
Gas-rich dust disks around young stars (hereafter, protoplanetary disks) provide the raw material to build up planets. Hence, it is critical to understand when and how they disperse. 
Observations of protoplanetary disks suggest that they follow a two-timescale evolution. In the first few Myr their dispersal is thought to be driven by viscous evolution, accretion of disk gas onto the central star. This stage is followed by a rapid ($\sim 10^5$\,yr) clearing attributed mainly to photoevaporation driven by the central star, which heats gas in the disk surface to thermal escape velocity (e.g. \citealt{alexander13} for a recent review). 

While observations of blueshifted lines tracing the disk surface demonstrate that photoevaporation is occurring in some systems (e.g. \citealt{ps09}), its role in clearing protoplanetary material is still debated. This is mainly because photoevaporative mass loss rates  are poorly constrained observationally. At the same time predicted values span two orders of magnitude for sun-like stars, from $\sim 10^{-10}$ to $\sim 10^{-8}$\,M$_\odot$/yr, depending on the high-energy photons dominating the photoevaporation \citep{alexander06,gh09,owen10}. 
The smallest  mass loss rate implies that photoevaporation contributes only to the latest stages of disk clearing, while the largest is close to the median mass accretion rate of $\sim$1\,Myr-old stars (e.g. \citealt{calvet00}) suggesting that photoevaporation can drive disk dispersal early on.

A critical input parameter to photoevaporative disk models is the amount of high-energy stellar radiation impinging on the disk. X-rays (0.1-10\,keV) from young stars are relatively well characterized and stellar X-ray luminosities are available for hundreds of stars in nearby star-forming regions (e.g. \citealt{getman05,preibisch05}). X-ray spectra typically peak around $\sim$1\,keV with only a few young stars sporting a large soft excess component at 0.3-0.4\,keV \citep{guedel07}\footnote{However, a soft excess, identified as anomalously high fluxes in lines forming at temperature of only a few MK, is seen in all accreting young stars (e.g. \citealt{gn09}).}. Thus, the stellar chromosphere and/or corona are thought to dominate their X-ray emission (e.g. \citealt{gn09} for a review). Hard X-rays (1-10\,keV) are also not easily absorbed by circumstellar matter (e.g. \citealt{ercolano09}), such as accretion columns or magnetically driven winds, hence measured stellar X-ray luminosities should provide a good estimate for the hard X-ray radiation reaching the disk surface. At the opposite side of the high-energy stellar spectrum, the H$_2$-dissociating far-ultraviolet
 (FUV; 6-13.6eV) luminosity of young sun-like stars has been found to be proportional to their accretion luminosity (e.g. \citealt{ingleby11,yang12}). This suggests that the FUV emission mostly traces disk gas accreting onto the star and shocked at the stellar surface. Finally, extreme-UV radiation (EUV; 13.6-100eV) is poorly constrained both for young (e.g. \citealt{alexander05}) as well as for old stars (e.g. \citealt{ribas05,linsky14}). This is because interstellar gas easily absorbs EUV photons thus hampering their direct detection. Similarly, stellar EUV photons may be absorbed in accretion columns or magnetically driven jets launched from the disk surface near the star (e.g. \citealt{alexander04,hg09}). Hence, the EUV radiation reaching the disk may be substantially lower than that emitted by the star. It is also debated if most of the EUV emission originates in the accretion shock or in the chromosphere  (e.g. \citealt{alexander04,herczeg07}). Only with a rather large chromospheric component ($>$10$^{41}$\,photons/s), which would not fade away as accretion declines, could EUV radiation shut off disk accretion and clear the protoplanetary material in a timescale consistent with that observed (e.g. \citealt{alexander06,aa09}).

Ground-based observations have demonstrated that the surface of protoplanetary disks can be ionized by the central star high-energy photons, e.g. via the detection and characteristic profile of the \neii{} emission line at 12.8\micron{} \citep{herczegetal07,ps09,najita09,baldovin12,sacco12}. In a previous contribution we showed that a fully or a partially ionized protoplanetary disk surface emits free-free cm radiation that 
should also be detectable with current astronomical facilities \citep{pascucci12}. We also derived analytic scaling relations between 
the ionizing radiation impinging on the disk and the free-free disk emission. Recent hydrodynamical model calculations of \citet{owen13} agree with our derived relations.
Hence, free-free cm emission can be used to constrain the high-energy radiation actually reaching the disk and photoionizing H. 

Building on these findings, here we provide stringent upper limits on the EUV photon luminosity reaching the disk for 14 young ($\sim 2-10$\,Myr) stars. 
The paper is organized as follows. In Sect.~\ref{sect:obs} we summarize new and nearly simultaneous cm observations of six young stars with disks obtained with the Australia Telescope Compact Array (ATCA). These six targets were chosen because they are relatively nearby, far away from massive stars, and have ancillary evidence of an ionized disk surface from the detection of the \neii{} line at 12.81\,\micron{} (see Table~\ref{tab:sou} for the main properties). We also preferred sources with no known jets because shocked ISM gas is known to produce \neii{} and free-free cm emission (e.g. \citealt{anglada98,van09}).  In Sect.~\ref{sect:results} we discuss the immediate results and show that all ATCA sources have cm emission in excess to the dust thermal emission. We then measure the excess cm emission for the ATCA sources as well as for eight other disks in the literature which have a good coverage at millimeter and centimeter wavelengths. Finally, in Sect.~\ref{sect:euv_limits} we assume that all excess cm emission is due to free-free disk emission and derive upper limits on the EUV photon luminosity impinging on the disk. 
We discuss the main implications of our findings in Sect.~\ref{sect:conclusions}.

\section{Observations and Data Reduction}\label{sect:obs}
Centimeter continuum observations were carried out with the 6$\times$22\,m antennas ATCA interferometer between 2012 October 18 and October 21.  We used the hybrid H214 array configuration, where five antennas are arranged with baselines between 82 and 247\,m, and the sixth antenna is on a 4.5\,km baseline. However, the atmospheric phase stability was too poor to calibrate the longest baselines therefore we discarded all the data from the sixth antenna.
Our observations were conducted with the Compact Array Broadband Backend (CABB), 2049 channels with a total bandwidth of 2\,GHz, dual sideband with frequency pairs centered at 33+35 GHz (8.8\,mm) and 17+19 GHz (17\,mm), and simultaneous observations at 9.0+5.5 GHz (3.3 and 5.5\,cm, respectively).
In addition to our six science targets we also observed the source 1934-638 for flux calibration and 1921-293 for bandpass calibration. For each target we also identified a nearby and bright gain/phase calibrator whose exposures were interleaved with the science target exposures (see Table~\ref{tab:log}).

The data reduction followed the standard CABB procedure described
in the ATCA user guide\footnote{http://www.narrabri.atnf.csiro.au/observing/users$\_$guide/users$\_$guide.html} and was carried out with the software
package MIRIAD version 1.5 \citep{sault95}. 
In brief, we checked individual exposures and flagged bad baselines, antennas, and/or times. 
For the 8.8 and 17\,mm data we used the option {\it opcor} in {\it atlod} and then {\it tsyscal=any} in {\it atfix}  to correct the fluxes for atmospheric opacity. We then used the task {\it mfcal} on the bandpass calibrator to determine bandpass corrections taking antenna 2 as reference antenna. The bandpass solution was transferred to the flux and gain/phase calibrators using {\it gpcopy}. The fluxes of the bandpass and gain/phase calibrators were then scaled to the absolute flux units using {\it gpboot} and {\it mfboot}. Finally, we copied the phase calibrations to the science targets with {\it gpcopy}.
For the 17\,mm data we preferred to use 1934-638 as bandpass calibrator because 1921-293 had large phase amplitudes (between +10$^{\circ}$ and -10$^{\circ}$) even after atmospheric opacity correction. At the end of the data reduction we measured the flux of 1921-293 at each frequency and could verify that it is within 10\% of the values reported in the ATCA webpage, except for the 19\,GHz band for which it is within 20\%. The difference at 19\,GHz may be due to poor weather conditions which might have also impacted the phase stability of 1921-293 at this frequency. Because we have not used 1921-293 to calibrate the 17+19\,GHz data, we assume a 10\% absolute flux calibration uncertainty at all frequencies investigated here. This value is also in agreement with previous analysis of ATCA data (e.g. \citealt{ubach12}).

To compute continuum flux densities, rms, and identify any extension beyond the synthesized beam, we Fourier transformed the complex visibility to produce images of the sky brightness distribution (for this task we used  {\it invert}, {\it clean}, and {\it restore} in MIRIAD). Cleaned ATCA maps using uniform weighting are shown in Figs.~\ref{fig:first3sources} and \ref{fig:next3sources}. The resulting fits images were loaded in the Common Astronomy Software Applications (CASA) package to measure flux densities or 3$\sigma$ upper limits.
For this last step, we first computed the rms in a polygonal area around the expected location of the target, which is given by the 2MASS coordinates (Table~\ref{tab:sou}). If the source is not detected we report an upper limit equal to 3 times the computed rms.
If a source is detected we provide the flux density within the 3\,rms closed contour except in a few instances where our procedure is clearly missing some significant flux\footnote{The flux density within the 2\,rms contour is larger than that within the 3\,rms plus the 10\% absolute flux calibration uncertainty}, hence we report the 2\,rms closed contour flux density. All flux densities and upper limits are listed in Table~\ref{tab:fluxes}.

\section{Immediate Results}\label{sect:results}
\begin{figure}[h]
	\centering
	\begin{tabular}{lll}

\begin{tikzpicture}
\tikzstyle{every node}=[font=\small]
\node[above right] (img) at (0,0) {\includegraphics[width=\plotwidth]{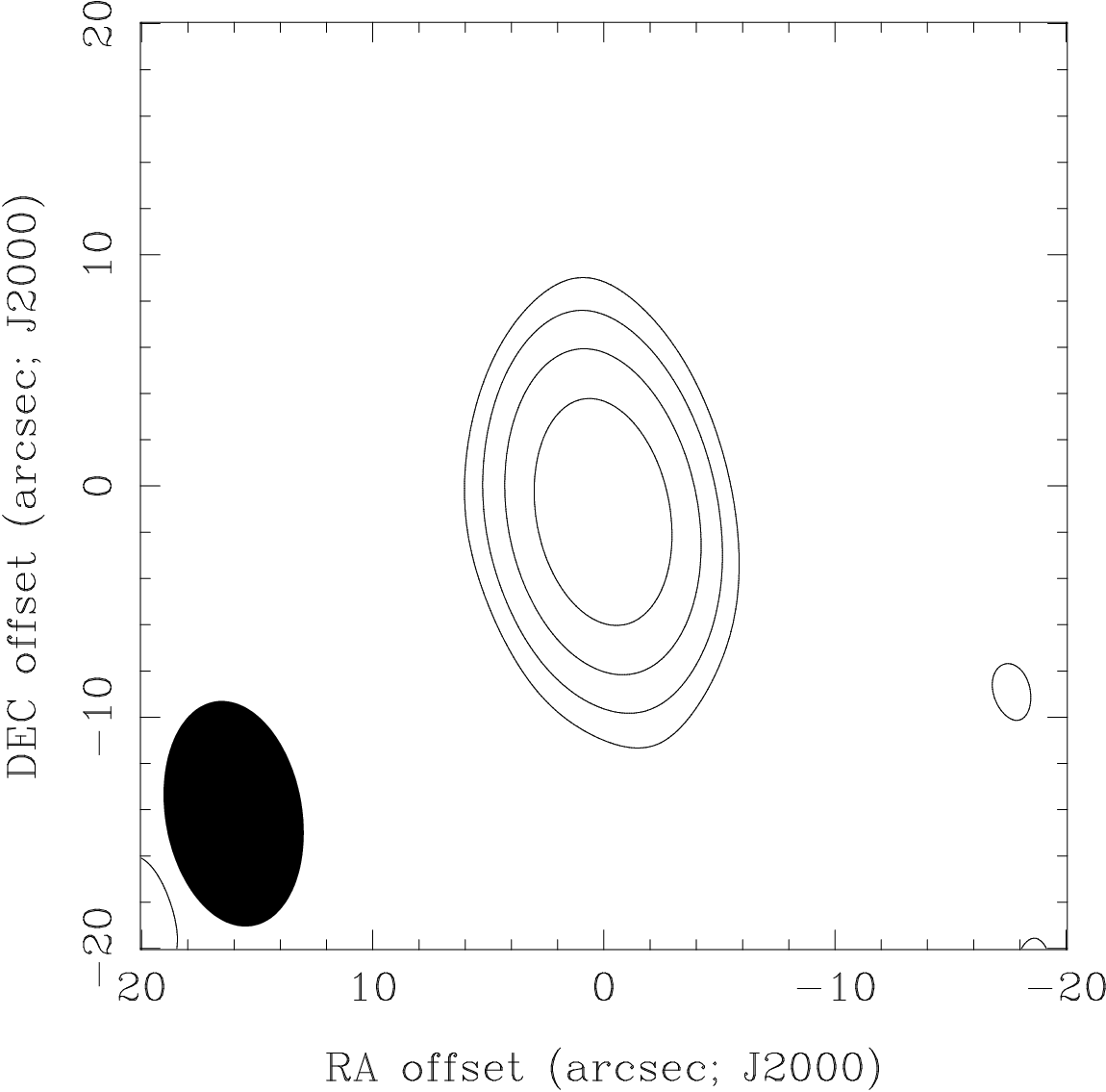}};
\pgfsize{\mywidth}{\myheight}
\node[text centered] at (\mywidth/2 + \widthoffset, \myheight*3/4 + \heightoffset) {V892 Tau  0.9cm};
\end{tikzpicture}
&

\begin{tikzpicture}
\tikzstyle{every node}=[font=\small]
\node[above right] (img) at (0,0) {\includegraphics[width=\plotwidth]{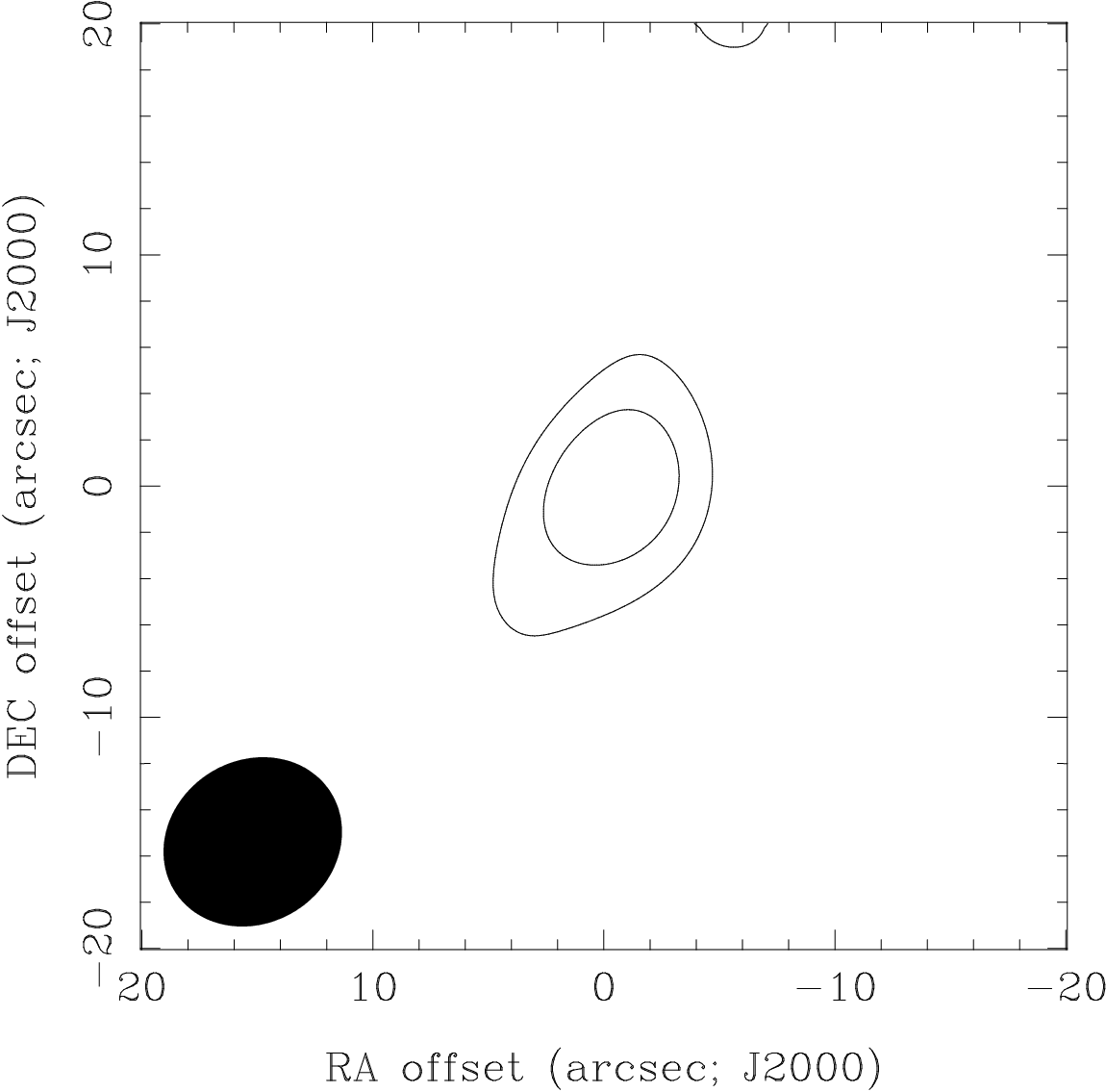}};
\pgfsize{\mywidth}{\myheight}
\node[text centered] at (\mywidth/2 + \widthoffset, \myheight*3/4 + \heightoffset) {SZ Cha  0.9cm};
\end{tikzpicture}
&

\begin{tikzpicture}
\tikzstyle{every node}=[font=\small]
\node[above right] (img) at (0,0) {\includegraphics[width=\plotwidth]{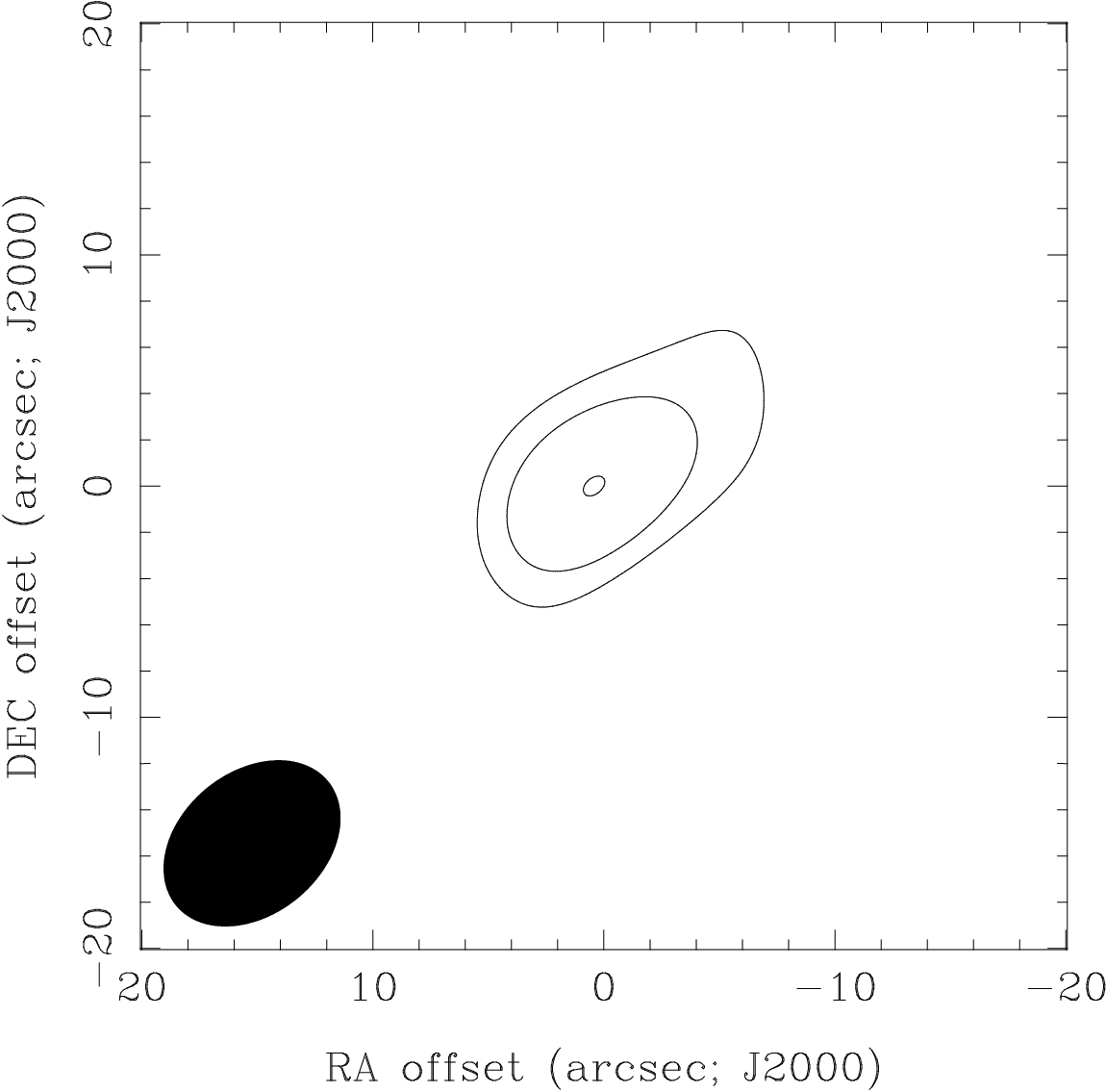}};
\pgfsize{\mywidth}{\myheight}
\node[text centered] at (\mywidth/2 + \widthoffset, \myheight*3/4 + \heightoffset) {CS Cha  0.9cm};
\end{tikzpicture}
\\

\begin{tikzpicture}
\tikzstyle{every node}=[font=\small]
\node[above right] (img) at (0,0) {\includegraphics[width=\plotwidth]{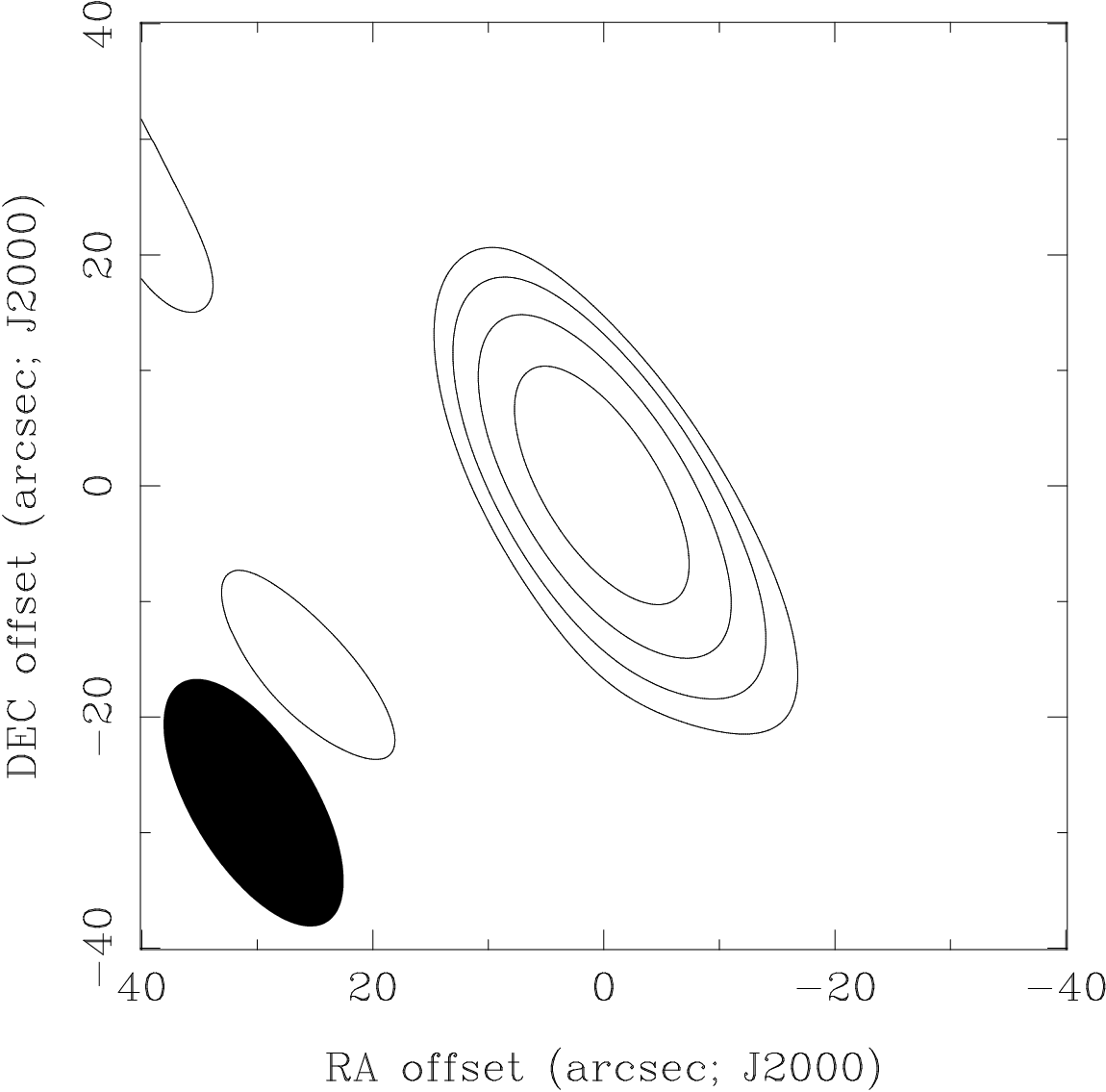}};
\pgfsize{\mywidth}{\myheight}
\node[text centered] at (\mywidth/2 + \widthoffset, \myheight*3/4 + \heightoffset) {V892 Tau  1.7cm};
\end{tikzpicture}
&

\begin{tikzpicture}
\tikzstyle{every node}=[font=\small]
\node[above right] (img) at (0,0) {\includegraphics[width=\plotwidth]{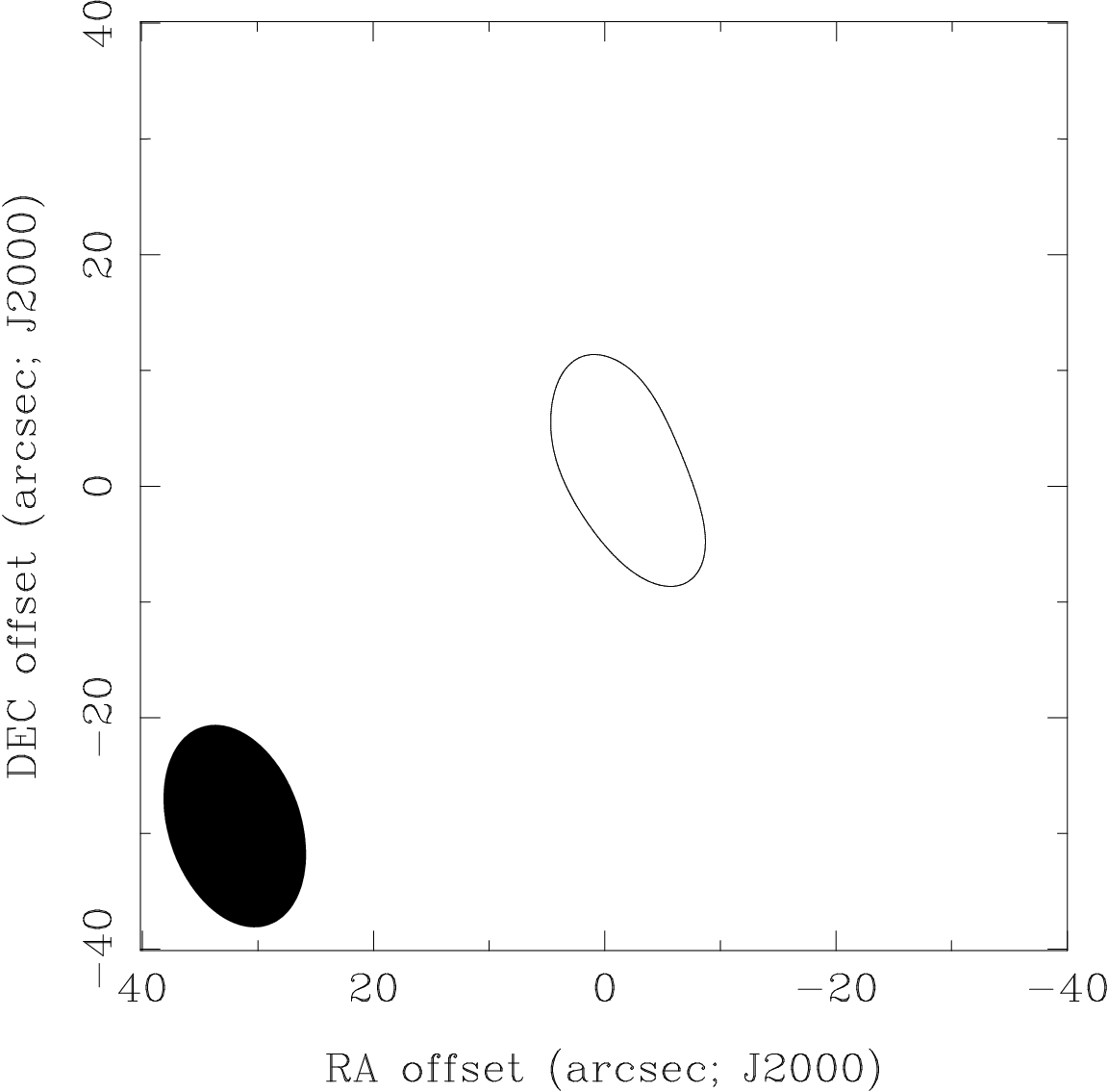}};
\pgfsize{\mywidth}{\myheight}
\node[text centered] at (\mywidth/2 + \widthoffset, \myheight*3/4 + \heightoffset) {SZ Cha  1.7cm};
\end{tikzpicture}
&

\begin{tikzpicture}
\tikzstyle{every node}=[font=\small]
\node[above right] (img) at (0,0) {\includegraphics[width=\plotwidth]{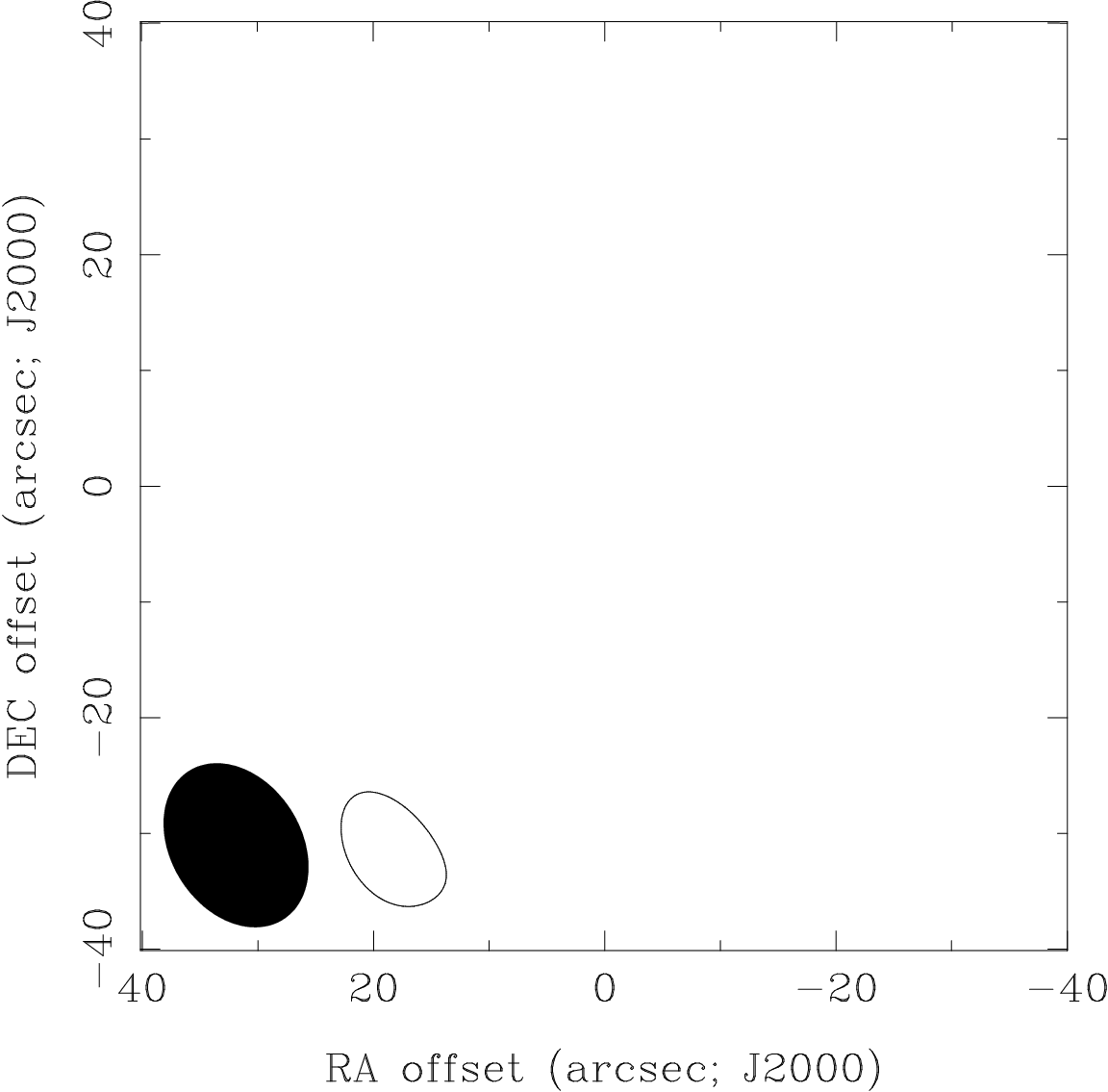}};
\pgfsize{\mywidth}{\myheight}
\node[text centered] at (\mywidth/2 + \widthoffset, \myheight*3/4 + \heightoffset) {CS Cha  1.7cm};
\end{tikzpicture}
\\

\begin{tikzpicture}
\tikzstyle{every node}=[font=\small]
\node[above right] (img) at (0,0) {\includegraphics[width=\plotwidth]{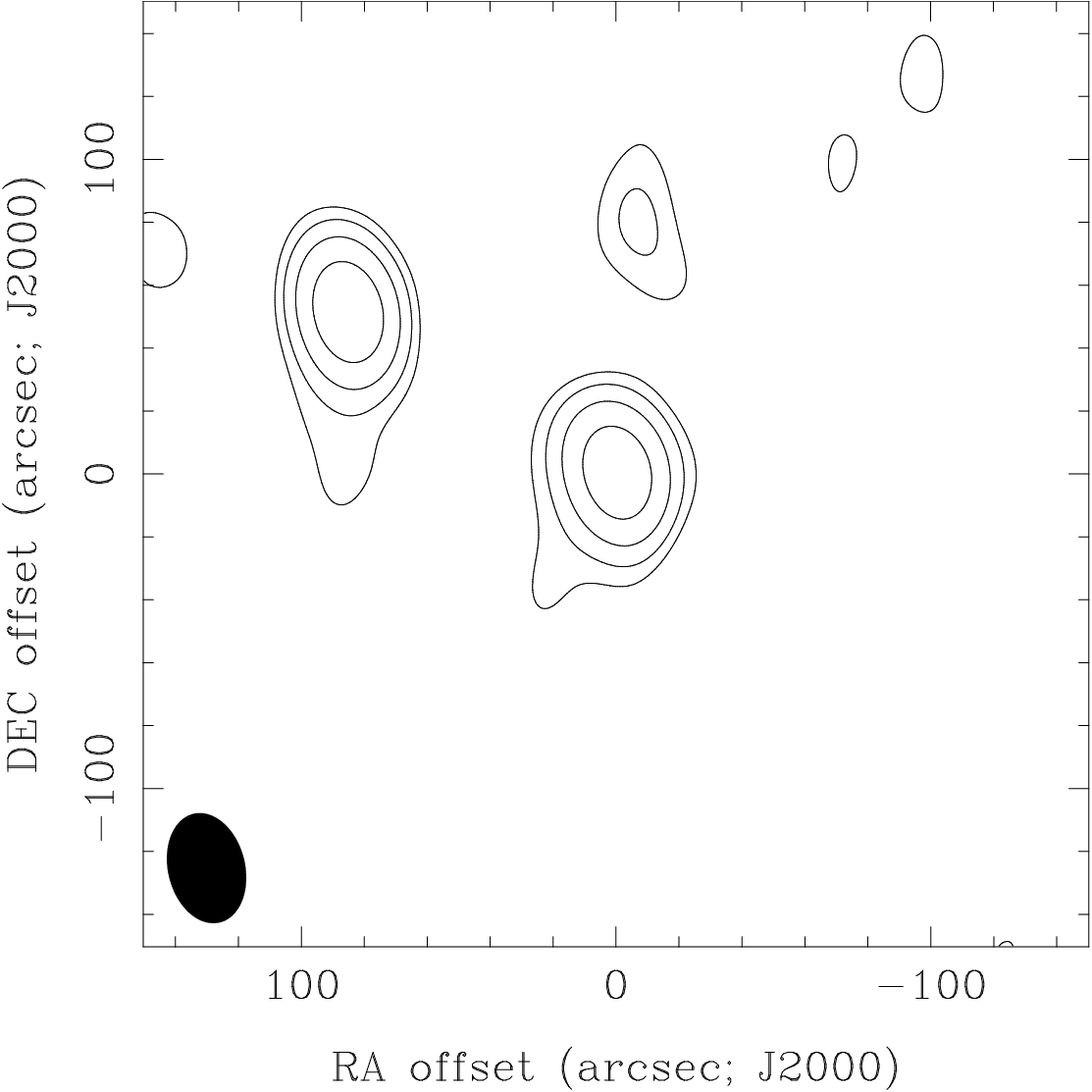}};
\pgfsize{\mywidth}{\myheight}
\node[text centered] at (\mywidth/2 + \widthoffset, \myheight*3/4 + \heightoffset) {V892 Tau  3.3cm};
\end{tikzpicture}
&

\begin{tikzpicture}
\tikzstyle{every node}=[font=\small]
\node[above right] (img) at (0,0) {\includegraphics[width=\plotwidth]{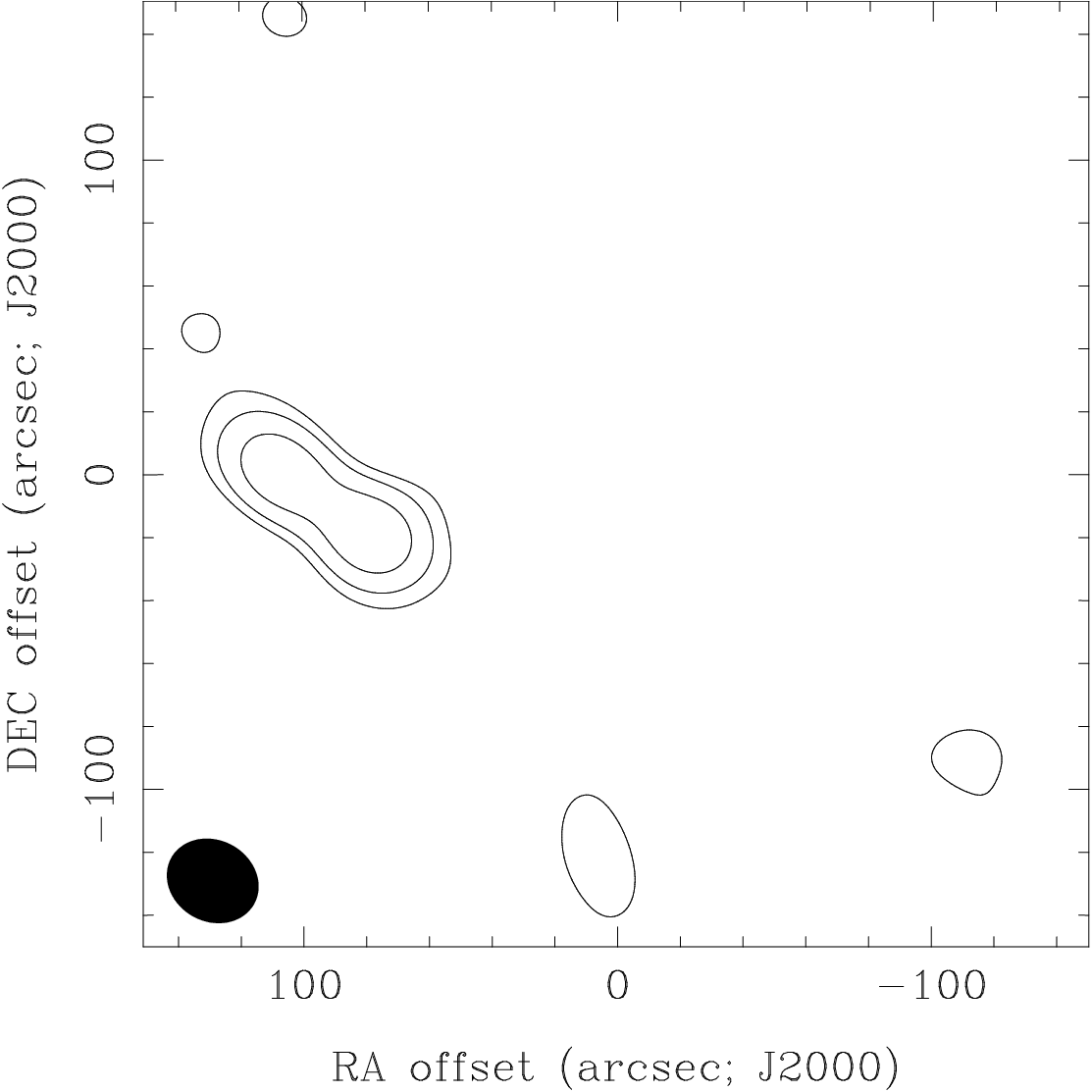}};
\pgfsize{\mywidth}{\myheight}
\node[text centered] at (\mywidth/2 + \widthoffset, \myheight*3/4 + \heightoffset) {SZ Cha  3.3cm};
\end{tikzpicture}
&

\begin{tikzpicture}
\tikzstyle{every node}=[font=\small]
\node[above right] (img) at (0,0) {\includegraphics[width=\plotwidth]{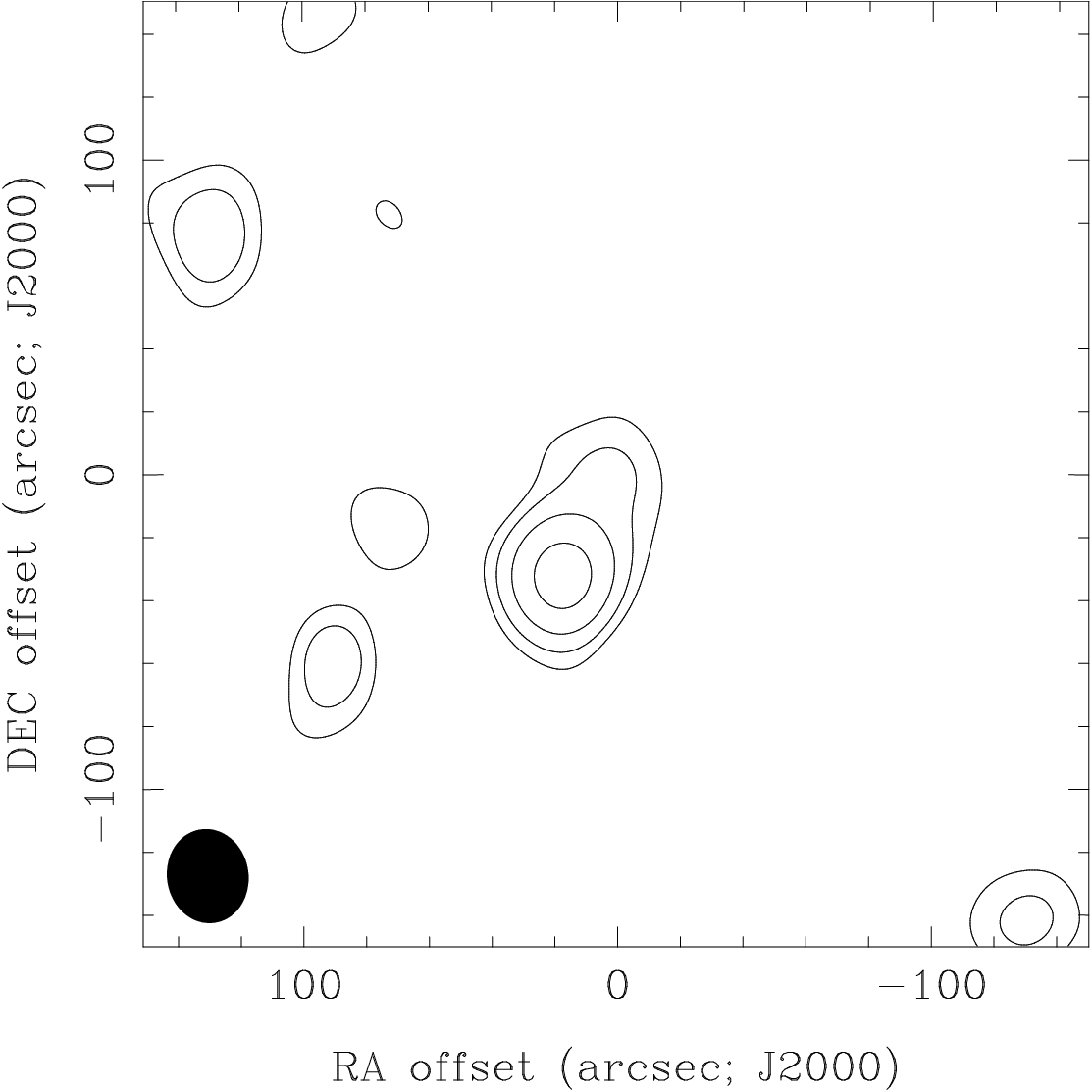}};
\pgfsize{\mywidth}{\myheight}
\node[text centered] at (\mywidth/2 + \widthoffset, \myheight*3/4 + \heightoffset) {CS Cha  3.3cm};
\end{tikzpicture}
\\

\begin{tikzpicture}
\tikzstyle{every node}=[font=\small]
\node[above right] (img) at (0,0) {\includegraphics[width=\plotwidth]{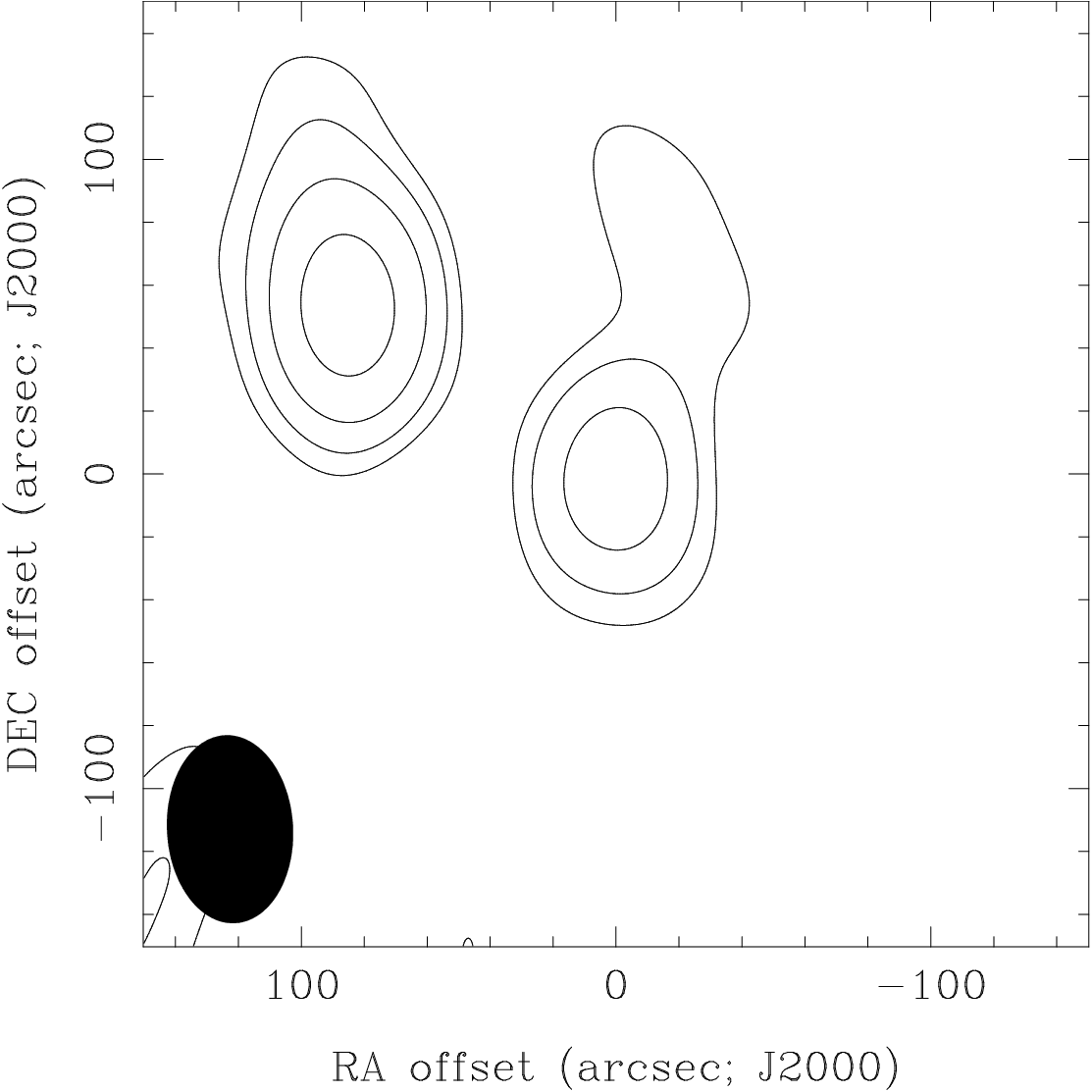}};
\pgfsize{\mywidth}{\myheight}
\node[text centered] at (\mywidth/2 + \widthoffset, \myheight*3/4 + \heightoffset) {V892 Tau  5.5cm};
\end{tikzpicture}
&

\begin{tikzpicture}
\tikzstyle{every node}=[font=\small]
\node[above right] (img) at (0,0) {\includegraphics[width=\plotwidth]{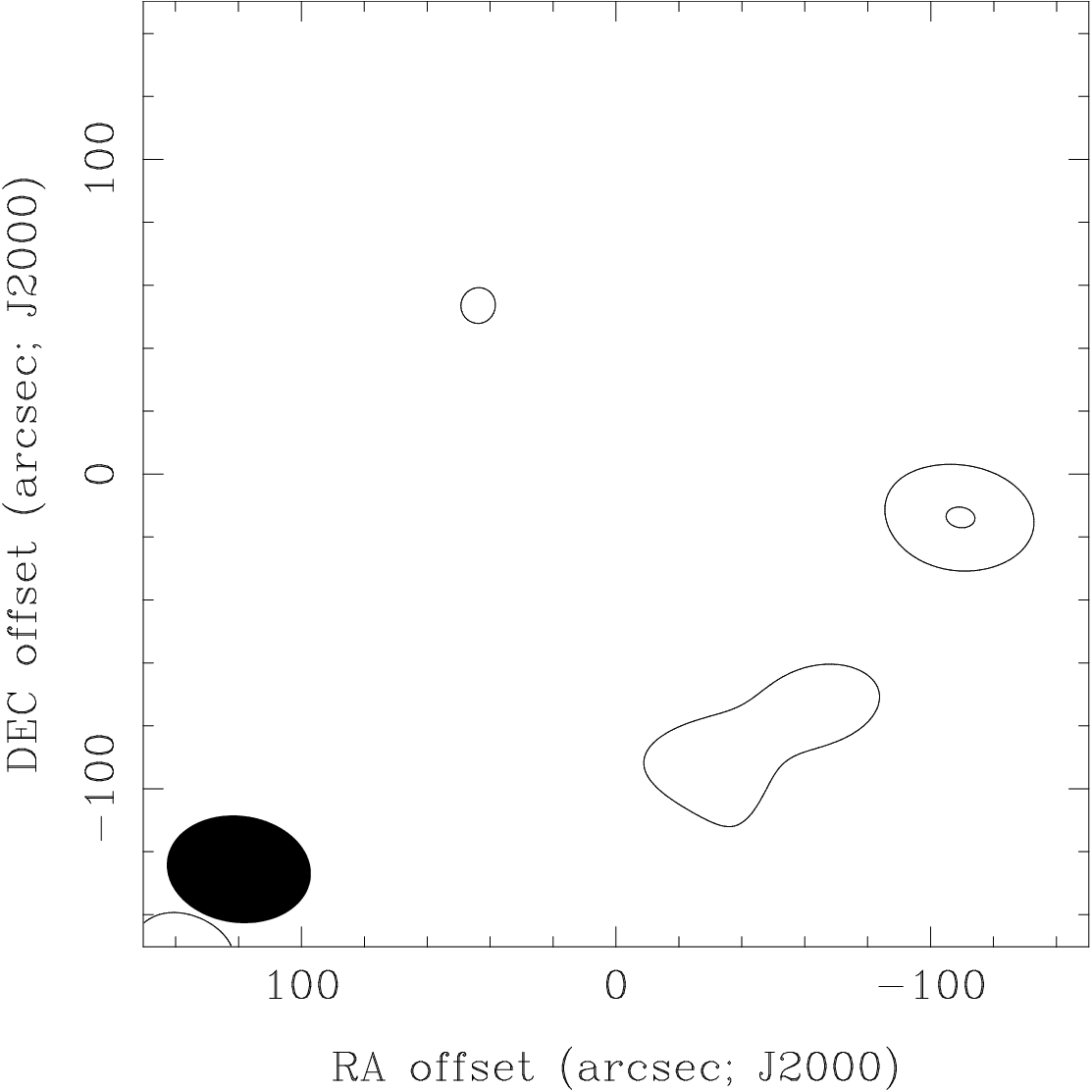}};
\pgfsize{\mywidth}{\myheight}
\node[text centered] at (\mywidth/2 + \widthoffset, \myheight*3/4 + \heightoffset) {SZ Cha  5.5cm};
\end{tikzpicture}
&

\begin{tikzpicture}
\tikzstyle{every node}=[font=\small]
\node[above right] (img) at (0,0) {\includegraphics[width=\plotwidth]{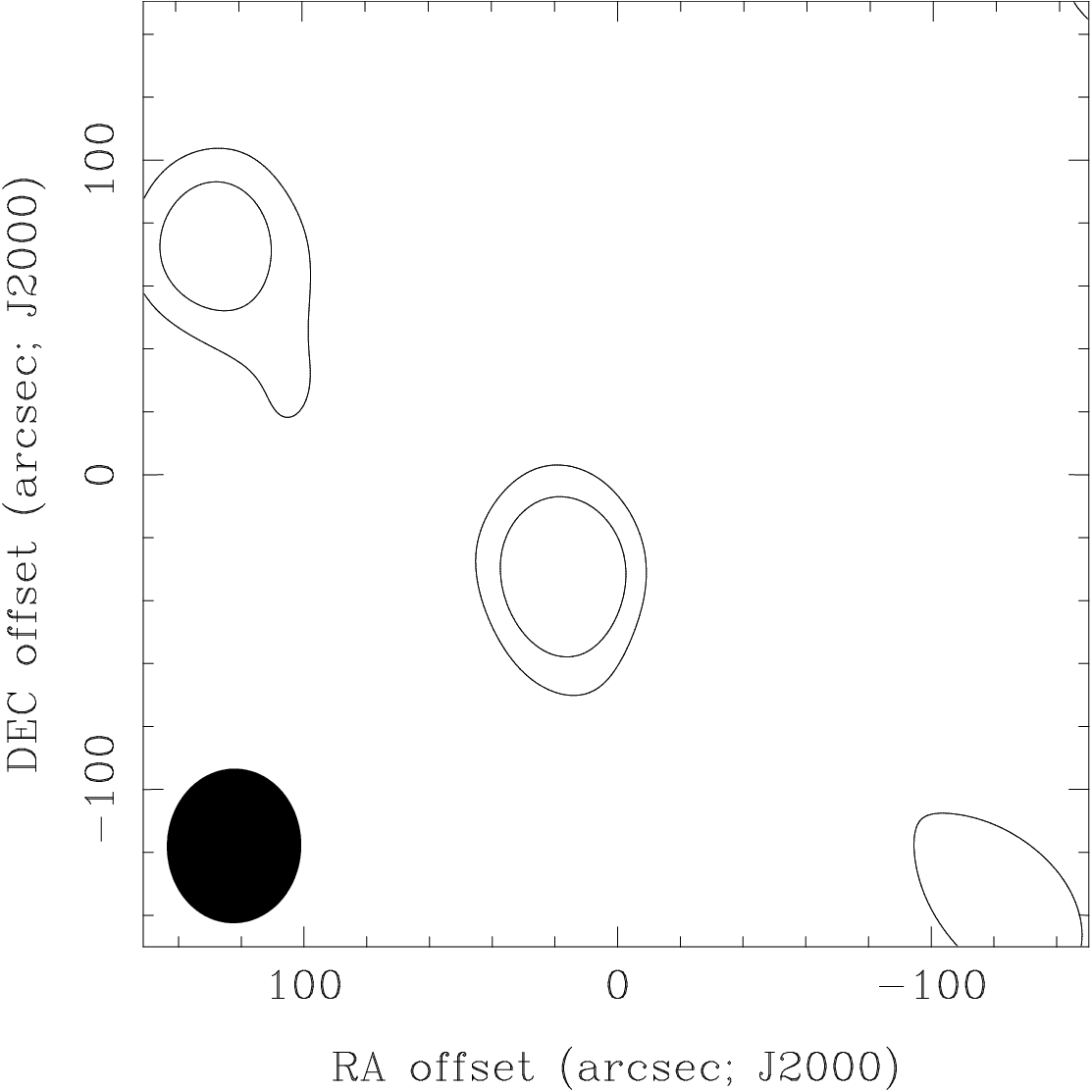}};
\pgfsize{\mywidth}{\myheight}
\node[text centered] at (\mywidth/2 + \widthoffset, \myheight*3/4 + \heightoffset) {CS Cha  5.5cm};
\end{tikzpicture}
\\

	\end{tabular}
	\caption{Cleaned ATCA maps using uniform weighting. In all panels the contours are 3, 6, 12, and 24 times the image rms. \label{fig:first3sources}}
\end{figure}

\begin{figure}[h]
	\centering
	\begin{tabular}{lll}

\begin{tikzpicture}
\tikzstyle{every node}=[font=\small]
\node[above right] (img) at (0,0) {\includegraphics[width=\plotwidth]{mpmus_7mm}};
\pgfsize{\mywidth}{\myheight}
\node[text centered] at (\mywidth/2 + \widthoffset, \myheight*3/4 + \heightoffset) {MP Mus  0.9cm};
\end{tikzpicture}
&

\begin{tikzpicture}
\tikzstyle{every node}=[font=\small]
\node[above right] (img) at (0,0) {\includegraphics[width=\plotwidth]{sr21_7mm}};
\pgfsize{\mywidth}{\myheight}
\node[text centered] at (\mywidth/2 + \widthoffset, \myheight*3/4 + \heightoffset) {SR 21  0.9cm};
\end{tikzpicture}
&

\begin{tikzpicture}
\tikzstyle{every node}=[font=\small]
\node[above right] (img) at (0,0) {\includegraphics[width=\plotwidth]{v4046sgr_7mm}};
\pgfsize{\mywidth}{\myheight}
\node[text centered] at (\mywidth/2 + \widthoffset, \myheight*3/4 + \heightoffset) {V4046 Sgr  0.9cm};
\end{tikzpicture}
\\

\begin{tikzpicture}
\tikzstyle{every node}=[font=\small]
\node[above right] (img) at (0,0) {\includegraphics[width=\plotwidth]{mpmus_1cm}};
\pgfsize{\mywidth}{\myheight}
\node[text centered] at (\mywidth/2 + \widthoffset, \myheight*3/4 + \heightoffset) {MP Mus  1.7cm};
\end{tikzpicture}
&

\begin{tikzpicture}
\tikzstyle{every node}=[font=\small]
\node[above right] (img) at (0,0) {\includegraphics[width=\plotwidth]{sr21_1cm}};
\pgfsize{\mywidth}{\myheight}
\node[text centered] at (\mywidth/2 + \widthoffset, \myheight*3/4 + \heightoffset) {SR 21  1.7cm};
\end{tikzpicture}
&

\begin{tikzpicture}
\tikzstyle{every node}=[font=\small]
\node[above right] (img) at (0,0) {\includegraphics[width=\plotwidth]{v4046sgr_1cm}};
\pgfsize{\mywidth}{\myheight}
\node[text centered] at (\mywidth/2 + \widthoffset, \myheight*3/4 + \heightoffset) {V4046 Sgr  1.7cm};
\end{tikzpicture}
\\

\begin{tikzpicture}
\tikzstyle{every node}=[font=\small]
\node[above right] (img) at (0,0) {\includegraphics[width=\plotwidth]{mpmus_3cm}};
\pgfsize{\mywidth}{\myheight}
\node[text centered] at (\mywidth/2 + \widthoffset, \myheight*3/4 + \heightoffset) {MP Mus  3.3cm};
\end{tikzpicture}
&

\begin{tikzpicture}
\tikzstyle{every node}=[font=\small]
\node[above right] (img) at (0,0) {\includegraphics[width=\plotwidth]{sr21_3cm}};
\pgfsize{\mywidth}{\myheight}
\node[text centered] at (\mywidth/2 + \widthoffset, \myheight*3/4 + \heightoffset) {SR 21  3.3cm};
\end{tikzpicture}
&

\begin{tikzpicture}
\tikzstyle{every node}=[font=\small]
\node[above right] (img) at (0,0) {\includegraphics[width=\plotwidth]{v4046sgr_3cm}};
\pgfsize{\mywidth}{\myheight}
\node[text centered] at (\mywidth/2 + \widthoffset, \myheight*3/4 + \heightoffset) {V4046 Sgr  3.3cm};
\end{tikzpicture}
\\

\begin{tikzpicture}
\tikzstyle{every node}=[font=\small]
\node[above right] (img) at (0,0) {\includegraphics[width=\plotwidth]{mpmus_6cm}};
\pgfsize{\mywidth}{\myheight}
\node[text centered] at (\mywidth/2 + \widthoffset, \myheight*3/4 + \heightoffset) {MP Mus  5.5cm};
\end{tikzpicture}
&

\begin{tikzpicture}
\tikzstyle{every node}=[font=\small]
\node[above right] (img) at (0,0) {\includegraphics[width=\plotwidth]{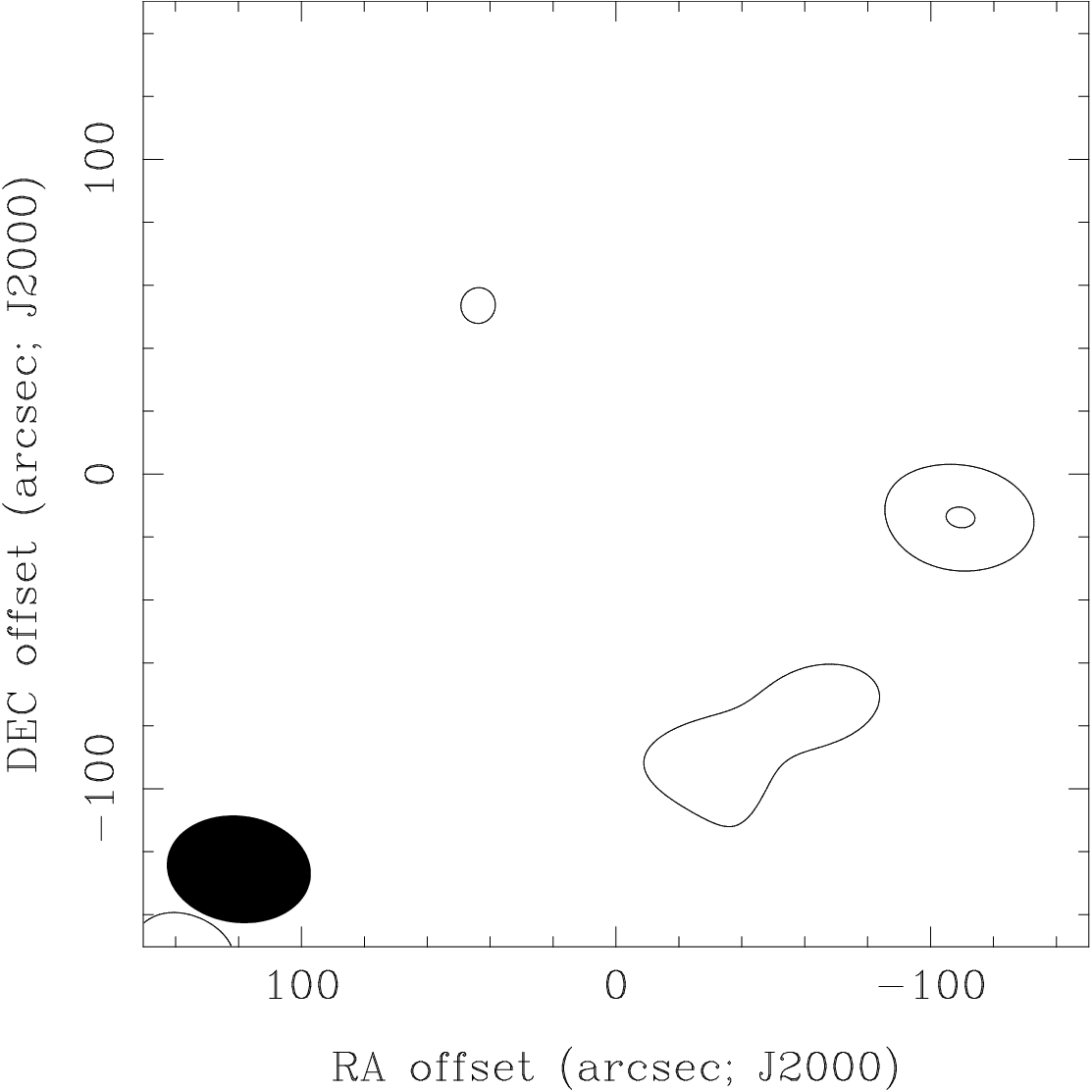}};
\pgfsize{\mywidth}{\myheight}
\node[text centered] at (\mywidth/2 + \widthoffset, \myheight*3/4 + \heightoffset) {SR 21  5.5cm};
\end{tikzpicture}
&

\begin{tikzpicture}
\tikzstyle{every node}=[font=\small]
\node[above right] (img) at (0,0) {\includegraphics[width=\plotwidth]{v4046sgr_6cm}};
\pgfsize{\mywidth}{\myheight}
\node[text centered] at (\mywidth/2 + \widthoffset, \myheight*3/4 + \heightoffset) {V4046 Sgr  5.5cm};
\end{tikzpicture}
\\

	\end{tabular}
	\caption{Cleaned ATCA maps using uniform weighting. In all panels the contours are 3, 6, 12, and 24 times the image rms.\label{fig:next3sources}}
\end{figure}

We detect all ATCA sources at 0.9\,cm, all but CS~Cha at 1.7\,cm, all but SZ~Cha at 3.3\,cm, and only V892~Tau and CS~Cha at 5.5\,cm (Figs.~\ref{fig:first3sources} and \ref{fig:next3sources}). 
Several other radio sources are detected in the larger areas covered at 3.3 and 5.5\,cm. The most relevant one for the interpretation of our data is Hubble~4, a young star in Taurus with no circumstellar disk but strong cm emission (see discussion in  Appendix~\ref{appendix:ATCAsources}).

To determine if sources are extended we fit their emission with a 2D gaussian (task {\it gaussfit} in CASA)  and compare the size of the gaussian with the restored beam FWHM reported in Table~\ref{tab:fluxes}.
With this approach we find that SZ~Cha and CS~Cha are marginally extended at 0.9\,cm (gaussian widths of 12$'' \times6''$ and 10$'' \times6''$ respectively), in agreement with \citet{ubach12} who report extended emission for these two sources at a slightly shorter (0.7\,cm) wavelength. SZ~Cha is also spatially resolved at 1.7\,cm (gaussian width of 29$'' \times13''$) while CS~Cha is clearly extended at 3.3\,cm (gaussian width of 35$'' \times26 ’’$, position angle$\sim$162$^{\circ}$). V892~Tau is only marginally resolved at 3.3\,cm but its emission is clearly extended at 5.5\,cm (see Fig.~\ref{fig:first3sources}).

Disk emission should be confined within the disk size, rarely larger than 1,000\,AU (e.g. \citealt{va05}). Hence, the extended emission at 3.3 and 5.5\,cm for V892~Tau and CS~Cha, which corresponds to sizes $\gtrsim$5,000\,AU, do not have a disk origin. In addition, for CS~Cha the peak emission at these long wavelengths is offset from the 2MASS source coordinates (see Fig.~\ref{fig:first3sources}).
High-resolution centimeter observations of young stellar objects have shown that some radio sources are elongated along a direction close to the axis of a jet/outflow detected at shorter wavelengths (see e.g. the case of HL~Tau in \citealt{rodmann06}). 
V892~Tau does not have any known jet/outflow \citep{kenyon08}. For CS~Cha, \citet{takami03} interpreted the positional displacement in the H$\alpha$ emission line as a micro-jet of $\sim$10\,mas in size and NE-SW direction, almost perpendicular to the elongation we see at 3.3\,cm. However, another possibility is that the H$\alpha$ displacement indicates the direction of the recently discovered stellar companion to CS~Cha (\citealt{guenther07} and Table~\ref{tab:sou}). Typical spectroastrometric displacement in the H$\alpha$ line for known binaries have angular scales of $\sim$0.2-0.4 times the binary separation \citep{takami03}, very close to the positional displacement measured for CS~Cha.
Although our targets were selected to be jet-free based on well established jet diagnostics such as the optical forbidden lines from O, N$^+$, and S$^+$ (e.g. \citealt{hartigan95}), jets/outflows remain the most plausible explanation for the extended cm emission seen in the ATCA images of these two sources. 
Based on these images the orientation of the jet is almost N-S for V892~Tau and NW-SE for CS~Cha. When the jet emission is not spatially resolved we report in Table~\ref{tab:fluxes} the flux density within the 3\,rms contour near the source and note that the emission is likely associated with a jet/outflow.

The presence of a jet is more difficult to ascertain for SZ~Cha. The marginal extensions at 7\,mm from \citet{ubach12} and at 0.9\,cm from our ATCA image indicate linear scales of $\sim$800-1,400\,AU. At the same time the extended 1.7\,cm emission is rather low S/N (see Table~\ref{tab:fluxes} and Fig.~\ref{fig:first3sources}) and the source is not detected at longer cm wavelengths where jet emission could become dominant (see the case of CS~Cha). One result that may point to the presence of a jet is the detection of a strong 
\neiii{} emission line at 15.55\,\micron{} resulting in an unusually high \neiii /\neii{} line flux ratio of $\sim$1 \citep{espaillat13}. This ratio suggests ionization by a rather hard EUV spectrum (L$_{\rm EUV} \propto \nu^{-1}$, \citealt{hg09}) impinging on the disk, which Espaillat et al. attribute to the central star. However, shocks can locally heat the ISM gas to very high temperatures ($>$10,000\,K) depending on their velocity and in a few cases are clearly associated with millions of K plasma also producing soft X-ray emission (see \citealt{frank14} for a recent review). It is possible that part or most of the \neiii{} emission toward SZ~Cha is not produced in the disk surface but in a jet  (see also the recent report of \neiii{} emission from the Sz~102 microjet, \citealt{liu14}). Additional observations are clearly required to establish if SZ~Cha is powering a jet, hence we will continue classifying this source as jet-free in our study.

\subsection{Excess Centimeter Emission}\label{sect:excess}
In this paper we call excess centimeter emission any long wavelength emission on top of the thermal dust disk emission. 
The first step in identifying excess cm emission is to 
assemble  the source spectral energy distribution (SED) and subtract off the thermal contribution from dust grains.

In assembling the SEDs of our ATCA sources we gather additional millimetric and centimetric fluxes from the literature. For V892~Tau we find millimeter fluxes in \citet{andrews05} and \citet{ricci12}. CS~Cha and SZ~Cha have been observed at 870\,\micron{} with APEX/LABOCA \citep{belloche11} and additional millimeter/centimeter fluxes are reported in \citet{ubach12}. MP~Mus has a 1.2\,mm detection from \citet{carpenter05} as well as 3\,mm and centimeter data in \citet{cortes09}. SR~21 has millimeter fluxes from various compilations (\citealt{am94}, \citealt{ricci10a}, \citealt{andrews11}, \citealt{ricci12}, \citealt{ubach12}). Finally, V4046~Sgr has been observed by \citet{jensen96}, \citet{rk10}, and \citet{oberg11} at millimeter wavelengths. A summary of literature fluxes and references is provided in Table~\ref{tab:litval}.

In addition to our six ATCA targets, we include in our study other young stellar objects from the literature that have: i) a good SED coverage at millimeter and centimeter wavelengths; and ii) stellar X-ray luminosities, since X-rays contribute to ionize the disk surface. These criteria result in eight additional sources: DG~Tau, DK~Cha, GM~Aur, HL~Tau, RY~Tau, TCha, TW~Hya, UZ~TauE. 
Source properties that are relevant to our study are summarized in Table~\ref{tab:litsou} while literature fluxes and references to assemble their SEDs are provided in Table~\ref{tab:litval}. All sources except DK~Cha have also infrared spectroscopy covering the \neii{} line at 12.81\,\micron{} (see Sect.~\ref{sect:low_wind}).

\begin{figure*}
\includegraphics[angle=0,width=\textwidth]{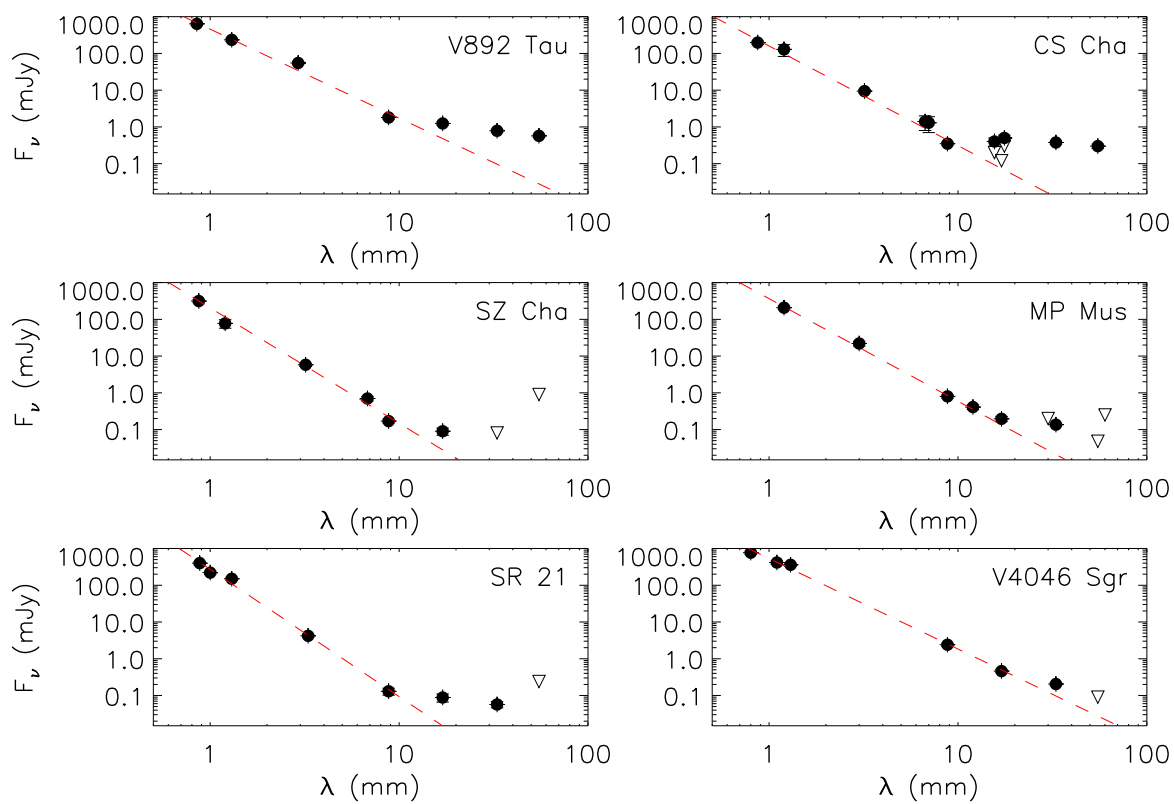}
\caption{SEDs of our targets with fluxes (filled circles) and upper limits (downward triangles) from this work and from the literature. In each panel the red dashed line is a linear fit to the millimeter fluxes between 0.8 and 10\,mm. These fits represent the contribution from the dust thermal emission. Note that for all sources there is an excess emission longward of 1\,cm.}
\label{fig:our}
\end{figure*}

\begin{figure*}
\includegraphics[angle=0,width=\textwidth]{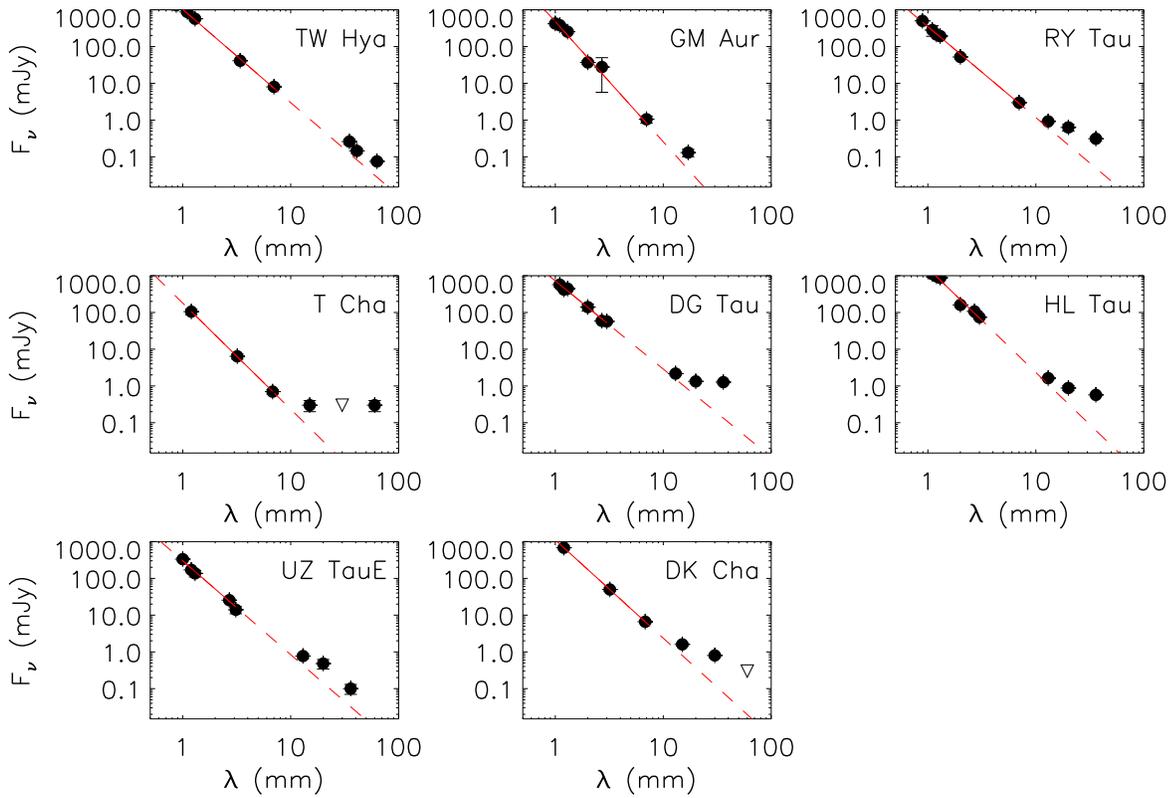}
\caption{Literature sources with good SED coverage at mm and cm wavelengths and known X-ray luminosities (see Sect.~\ref{sect:excess}). Symbols are as in Fig.~\ref{fig:our}.}
\label{fig:lit}
\end{figure*}

Figs.~\ref{fig:our} and \ref{fig:lit} show the long wavelength portion of the SED of our ATCA targets and literature sources.
Because thermal dust emission is mostly optically thin at millimeter wavelengths, the flux density can be written as $F_\nu \propto \nu^{\alpha_{\rm mm}}$. In fitting this relation we set a minimum uncertainty of 10\% for each flux density.
The fit to the millimeter fluxes\footnote{The longest wavelength we include in our fit is 10\,mm.}, ignoring the cm excess emphasized here, results in slopes ranging from 2.4 to a maximum value of 3.5 for SR~21, with uncertainties between $\sim$0.05 and 0.1. This range in $\alpha_{\rm mm}$ is similar to that found in nearby star-forming regions (e.g. \citealt{ricci10b,ubach12}). The important result is that for none of the sources studied here can thermal dust emission account for the measured centimeter fluxes.

Several physical mechanisms are known to produce radio emission. In Sect.~\ref{sect:results} we discussed two ATCA sources with clearly extended cm emission and concluded that most of the emission is likely coming from shocked gas in jets.
This gas can be ionized and produce cm free-free emission (e.g. \citealt{anglada98}). Similarly, a fully or partially ionized disk surface emits free-free cm radiation (e.g. \citealt{pascucci12}). Non-thermal (gyrosynchrotron) cm emission originating in magnetic fields has been also detected in late type dwarfs and several classes of active stars (e.g. \citealt{guedel02}). Finally, a population of  very large (cm-size) grains can produce extra cm thermal emission (e.g. \citealt{wilner05}) while very small (nm-size) spinning grains produce electric dipole emission in the microwave range \citep{rafikov06}. These different mechanisms can, and likely do, operate concurrently in young accreting stars surrounded by disks.  Multi-epoch and multi-wavelength radio observations can be used to assess the dominant physical mechanism (see Sect.~\ref{sect:radioslopes}) but the contribution of each process cannot be quantified with certainty. Therefore, in Sect.~\ref{sect:euv_limits} we will assume that all excess cm emission is due to the ionized disk surface and thus compute  
upper limits on the stellar EUV luminosity reaching the disk. We will show that even upper limits place interesting constraints on disk dispersal theories.

\subsection{Radio Spectral Indices}\label{sect:radioslopes}
Thermal bremsstrahlung (free-free) radiation and non-thermal gyrosynchroton emission are characterized by power-law spectra of the form $F_\nu \propto\nu^{\alpha_{\rm cm}}$. Cm- and nm-dust emission have a more bell-like spectral shape, which however cannot be easily recognized with two-three data points typically available at cm wavelengths. Therefore, we follow common practice and investigate the origin of the cm emission assuming that it has a power-law spectrum and thus compute the radio spectral index $\alpha_{\rm cm}$.

It is worth pointing out that a large range of spectral indices is expected both for free-free and gyrosynchroton emission. In the case of free-free emission $\alpha_{\rm cm}$ can be as low as -0.1 for optically thin emission (e.g. for an ionized disk surface, \citealt{pascucci12}) and up to +2 for optically thick emission. Values of $\sim$0.5 are predicted by \citet{owen13} for a disk surface heated and ionized by stellar X-rays. The spectral index for gyrosynchrotron emission depends on the energy distribution of the electrons and can vary between -2 and +2 (e.g. the review on stellar radio properties by \citealt{guedel02}). 

\begin{figure*}
\includegraphics[angle=0,width=\textwidth]{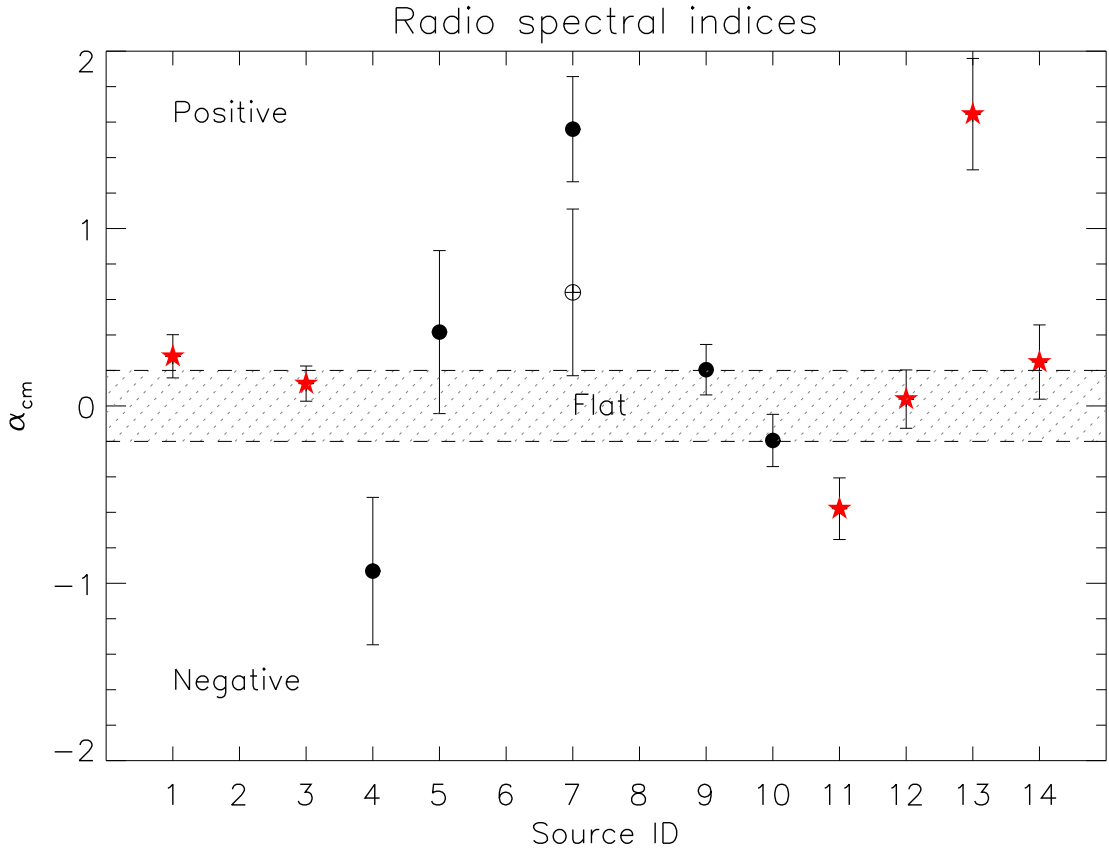}
\caption{Radio spectral slopes for sources with more than one excess cm flux measurement. Note that Sz~Cha (ID~2), V4046~Sgr (ID~6), and GM~Aur (ID~8) have detected cm excess emission only at one wavelength, hence they are not included in the figure. Red stars are sources with known jets. For TW~Hya (ID~7) we also compute the spectral slope using only the two longest wavelengths at 4.1 and 6.3\,cm (empty circle), see discussion in Sect.~\ref{sect:radioslopes}. Flat sources are consistent with free-free emission from optically thin, as from an ionized disk surface, to slightly thick plasma. Negative slopes
are suggestive of gyro-synchrotron emission.  The very positive slope of source
13 points to optically thick free-free jet emission.}
\label{fig:radioslopes}
\end{figure*}

Recently, \citet{dzib13} published a deep 4 and 6\,cm VLA survey of the Ophiucus star-forming complex and detected 56 known young stellar objects: Class~O/I (protostars), Class~II (accreting stars with disks), and Class~III objects (non-accreting stars). They find a large range of $\alpha_{\rm cm}$ in each sub-class but average values that decrease with evolutionary stage, from $\sim 0.5$ for the Class0/I to 0 for Class~II sources to -0.4 for Class~III sources (their Fig.~3).  At the same time, they see that the average variability (over a month timescale) and flux density increase with evolutionary stage (their Figs.~4 and 5). They interpret these trends as free-free emission dominating in Class~0/I and Class~II sources (and getting more optically thin in the Class~II stage) and gyrosynchroton emission dominating in Class~III sources. We follow the approach of \citet{dzib13} in identifying three classes of spectral indices: Positive ($\alpha_{\rm cm}>+0.2$), Flat ($-0.2 \ge \alpha_{\rm cm} \le +0.2$), and Negative ($\alpha_{\rm cm}<-0.2$). However, in computing the spectral indices we subtract off the contribution from thermal dust emission because only a few of our sources have 6\,cm detections, where the dust contribution is likely negligible.  Fig.~\ref{fig:radioslopes} shows the radio spectral indices for those sources that have at least two detections of excess cm emission, thus excluding Sz~Cha (ID~2), V4046~Sgr (ID~6), and GM~Aur (ID~8). The uncertainties in $\alpha_{\rm cm}$ include the uncertainty on the slope of the dust thermal emission. 

Seven of our sources have flat $\alpha_{\rm cm}$, consistent with free-free emission from optically thin to moderately thick plasma, thus including jet emission. V892~Tau and CS~Cha (IDs 1 and 3), whose ATCA cm emission is extended and most likely dominated by jet emission (Sect.~\ref{sect:results}), fall in this category. Only three sources (SR~21-ID~5, RY~Tau-ID~9, and T~Cha-ID~10) have flat $\alpha_{\rm cm}$ and no evidence of jets. Of them, only SR~21 and T~Cha have spectral slopes consistent with optically thin free-free disk emission while  emission from RY~Tau may be moderately thick as expected in the X-ray irradiated disk model of \citet{owen13}. Two sources, MP~Mus and DG~Tau have negative spectral indices. \citet{dzib13} also find Class~II sources in Ophiucus with negative spectral indices, indicating that substantial contribution from gyrosynchroton emission can be present in some disk sources. Multi-epoch cm observations would be helpful to further assess the non-thermal origin of the cm emission in these sources.

Finally, we wish to discuss the case of TW~Hya, which sports one of the largest positive $\alpha_{\rm cm}$ in our sample and has no evidence of a jet (ID~7, filled circle in Fig.~\ref{fig:radioslopes}). This  $\alpha_{\rm cm}$ is computed from the excess emission at 3 wavelengths: 3.5, 4.1, and 6.3\,cm. The latter two points are new VLA band-integrated flux densities from \citet{menu14}. These authors note that the spectral slope in the 1-GHz-band\footnote{these flux densities are not published} centered at 4.1\,cm is different from that at 6.3\,cm, it becomes flatter at the longer wavelength. They interpret this result as an indication of dust still contributing to the 4.1\,cm emission. We have thus re-computed  $\alpha_{\rm cm}$ neglecting the 3.5\,cm flux density, where dust would contribute even more to the emission. Indeed, we find that the new $\alpha_{\rm cm}$ (empty circle in Fig.~\ref{fig:radioslopes}) is substantially reduced, confirming the interpretation of \citet{menu14}. Thus, even this new  $\alpha_{\rm cm}$ should be considered an upper limit and it is likely that TW~Hya has also a flat cm spectral index. Sensitive observations at wavelengths longer than 6\,cm are needed to test this hypothesis.

Although in some cases $\alpha_{\rm cm}$ is not consistent with free-free disk emission, 
we will nevertheless assume that all the excess cm emission is from free-free so that we can derive upper limits to the EUV luminosity impinging on the disk (Sect.~\ref{sect:PhiEUV}).

\section{Ionizing Radiation Reaching the Disk}\label{sect:euv_limits}
A necessary condition to estimate the stellar ionizing luminosity reaching the disk atmosphere and ionizing the disk surface is that the associated free-free emission is optically thin. 
In \citet{pascucci12} we showed that even a partially ionized wind at 5,000\,K becomes optically thick at wavelengths longer than 20\,cm for plausible wind values. We can also compute the EUV photon luminosity above which the free-free emission would become optically thick using the expression for the continuum optical depth in \citet{bs78} and relating it to the emission measure as in eq.~5 from \citet{hg09}. In doing this calculation we assume a fully ionized region at 10,000\,K, as appropriate for the EUV case, with an emitting radius equal to the gravitational radius\footnote{The gravitational radius is where the hydrogen thermal speed is equal to the escape speed from the star gravitational field, e.g. \citet{hollenbach94}}, since most of the emission measure comes from regions close to this radius. As done in previous papers (e.g. \citealt{hg09}) we also assume that the fraction of the stellar photons intercepted by the disk is 0.7 and restrict ourselves to solar-mass stars\footnote{The value 0.7 corresponds to a disk vertical extent z$_{\rm max}$ = r where r is the midplane radial distance from the star. This is the vertical extent appropriate for EUV ionization.}. With this approach we find that the 3.3 and 5.5\,cm free-free emission becomes optically thick for EUV luminosities  
$\ge 3\times 10^{43}$\,s$^{-1}$ and $\ge 9\times 10^{42}$\,s$^{-1}$ respectively. We will see shortly that the upper limits we derive from the excess cm emission are lower than these values, meaning that free-free disk emission is optically thin at these wavelengths and we can use it to constrain the EUV luminosity impinging on the disk.
We also note that neglecting the absorption of EUV photons by dust grains is justified. In fact, ISM dust provides an optical depth to Lyman continuum photons greater than one in the ionized surface zone only for $\ge 10^{45}$\,s$^{-1}$ (eq.~6.3 in \citealt{hollenbach94}), well above the luminosities that we will derive here. In addition, because of dust growth and settling in disks (e.g. \citealt{testi14}), the opacity of the dust at the disk surface will be reduced with respect to the ISM value, further increasing the luminosity above which dust significantly absorbs EUV photons.

\subsection{Upper Limits on the EUV Luminosity}\label{sect:PhiEUV}
 Since several physical processes can produce radio emission (see Sects.~\ref{sect:results} and \ref{sect:radioslopes}), by assuming that all excess cm emission is due to free-free disk emission we will obtain upper limits to the ionizing radiation reaching the disk. Both EUV and X-rays can photoionize H atoms. In \citet{pascucci12} we used a gas temperature of 5,000\,K to estimate the free-free contribution from soft X-rays for the nearby disk of TW~Hya, a source with an exceptionally soft X-ray spectrum (e.g. \citealt{kastner02}). While hard X-rays heat the gas at lower temperatures thereby producing less free-free cm emission \citep{pascucci12}, the relative contribution of soft- vs hard-X-rays depends on the source X-ray spectrum, which is not always well characterized. Hence, we prefer to provide here conservative upper limits on the EUV radiation by not subtracting off the X-ray free-free contribution. Hereafter, we will use the notation $\Phi_{\rm EUV,cm}$ to refer to these upper limits derived from the excess cm emission. 


To estimate $\Phi_{\rm EUV,cm}$ we proceed as follows. First, we fit the millimeter fluxes to calculate the contribution from dust thermal emission at cm wavelengths as discussed in Sect.~\ref{sect:results}. Next, for each cm wavelength where we detect excess emission we estimate $\Phi_{\rm EUV,cm}$ using a generalization of eq.~2 in \cite{pascucci12} from the measured cm fluxes minus the dust thermal emission: the excess cm emission is directly proportional to the EUV luminosity reaching the disk \citep{pascucci12,owen13}. The main assumptions here are that the characteristic temperature of the gas is 10,000\,K and the 
fraction of photons absorbed by the disk is 0.7. This factor accounts only for the disk geometry, a further reduction in the ionizing radiation can occur because of absorption in the circumstellar matter, e.g. accretion columns and/or magnetically driven winds.

The most stringent upper limits on the EUV luminosity reaching the disk are summarized in Table~\ref{tab:euv_inferred} together with the wavelength providing such limits. The most sensitive wavelengths to place such upper limits are typically around 3 and 6\,cm, see last column of Table~\ref{tab:euv_inferred}. Note that the new 6.3\,cm flux density for TW~Hya reduces the upper limit on $\Phi_{\rm EUV,cm}$ by a factor of $\sim$3 with respect to what we obtained from the 3.5\,cm datapoint \citep{pascucci12} and that the $\Phi_{\rm EUV,cm}$ for GM~Aur is consistent with that reported by \citet{owen13}.
The main uncertainty in these upper limits comes from the uncertainty associated with the slope of the dust thermal emission ($\alpha_{\rm mm}$). Thus, we have also computed $\Phi_{\rm EUV,cm}$ assuming a steeper dust SED with  $\alpha_{\rm mm}$ minus the 1$\sigma$ uncertainty on the dust spectral slope. We find that upper limits derived at the longest wavelength ($\sim$6\,cm) typically increase from a few up to several \% (less than the absolute flux calibration uncertainty), because at these wavelengths dust emission contributes little. At wavelengths close to 1.5\,cm changes in $\Phi_{\rm EUV,cm}$ can be up to a factor of 2. Thus, in discussing our results we will assume that the  $\Phi_{\rm EUV,cm}$ derived with our approach can be at most off by a factor of 2. In other words, the upper limits could be at most a factor of 2 higher than those provided in Table~\ref{tab:euv_inferred}.

\begin{figure*}
\includegraphics[angle=0,width=\textwidth]{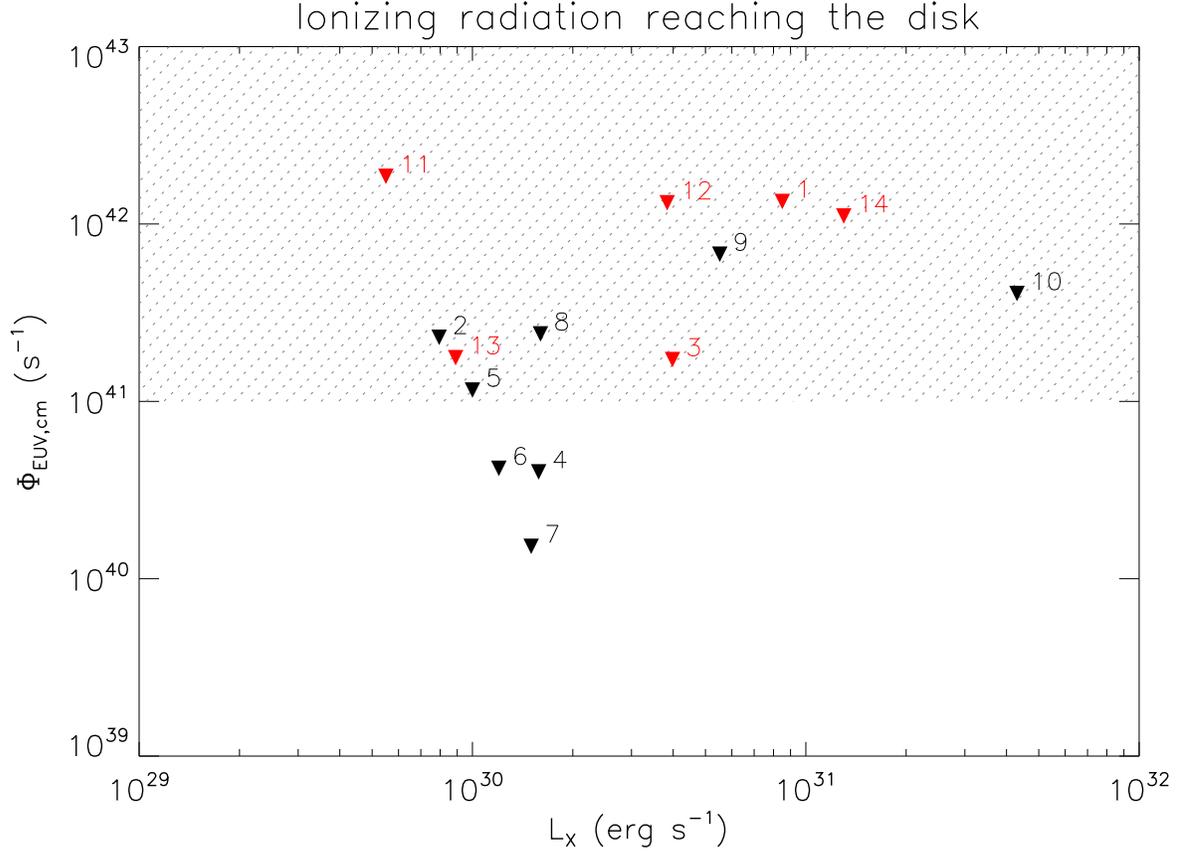}
\caption{Upper limits to the EUV photon luminosity reaching the disk ($\Phi_{\rm EUV,cm}$) as a function of stellar X-ray luminosity (L$_{\rm X}$). $\Phi_{\rm EUV,cm}$ is estimated from the excess cm emission. 
Sources with known jets are marked in red. The grey area shows the range of stellar EUV luminosities derived by \citet{alexander05}. Assuming an EUV photon energy of 13.6\,eV the conversion factor from photons/s to erg/s is $\sim 2.2 \times 10^{-11}$, meaning that an EUV luminosity of $10^{41}$\,s$^{-1}$ corresponds to $2.2 \times 10^{30}$\,erg/s.}
\label{fig:euv}
\end{figure*}

We find a broad range of $\Phi_{\rm EUV,cm}$ from  $\sim 2 \times 10^{40}$ to $10^{42}$\,s$^{-1}$ with no obvious correlation with the stellar X-ray luminosity (Fig.~\ref{fig:euv}). Sources with known jets/outflows (red symbols in Fig.~\ref{fig:euv}) have higher $\Phi_{\rm EUV,cm}$ than those without, on average by an order of magnitude. In these sources most of the cm emission is likely arising from shocked ISM material and does not trace the ionized disk surface or the ionizing luminosity from the star. This is confirmed in a few instances, such as in CS~Cha, where the cm emission is found to be spatially extended and does not peak at the stellar location (see discussion in Sect.~\ref{sect:results}). If we exclude the sources with jets and average the other upper limits we find that $\Phi_{\rm EUV,cm}$ is at most $2\times10^{41}$\,s$^{-1}$. 

How do these upper limits compare with the ionizing radiation emitted by the star? 
\citet{ribas05} used a small sample of solar analogs with ages between $\sim0.1$-7\,Gyr and could show that there is a tight correlation between the 1-1,200\,\AA{} stellar flux scaled at 1\,AU and stellar age (their Fig.~6 and eq.~1). 
Using their relation for the EUV 100-920\,\AA{} interval we find a luminosity of $\sim 3 \times 10^{40}$\,s$^{-1}$ for solar analogs that are 100\,Myr old. This luminosity is close to the upper limits we estimate for the three $\sim$5-10\,Myr-old sources in our sample, namely MP~Mus, V4046~Sgr, and TW~Hya. We do not know whether the Ribas et al. relation holds for sources younger than 100\,Myr but 
based on X-ray and FUV studies \citep{ingleby11} we should expect an increase in the stellar EUV luminosity at least back to 10\,Myr. The same studies have also shown that the stellar X-ray emission remains rather constant over the 1-10\,Myr age range we investigate here, while FUV radiation decreases in the same time interval and reaches the chromospheric level of $L_{\rm FUV}/L_{\rm star}$ of $\sim 10^{-4}$ in non-accreting stars. The behavior of the EUV emission in the 1-10\,Myr age range will depend on whether it mostly traces accretion (as FUV, in which case it should decrease with time) or the chromosphere (as X-ray, in which case it could be flat).

In the pre-main sequence regime, \citet{alexander05} estimated order-of-magnitude EUV luminosities (between 700-912\,\AA{}) by modeling literature emission measures from five sun-like stars that are a few Myr old. They find a broad range of luminosities ($10^{41}-10^{44}$\,s$^{-1}$) which we show as a dashed region in our Fig.~\ref{fig:euv}. For the only source we have in common, RY~Tau, our EUV upper limit reaching the disk is only a factor of $\sim$3 lower than theirs, basically consistent with their measurement given the uncertainties on these values. However, in general our $\Phi_{\rm EUV,cm}$ lie on the lowest side of the stellar ionizing luminosities they infer. In addition, the $\Phi_{\rm EUV,cm}$ for the $\sim$5-10\,Myr-old systems TW~Hya, MP~Mus, and V4046~Sgr is several times lower than $10^{41}$\,s$^{-1}$. \citet{herczeg07} estimated an ionizing photon luminosity of $\sim 5 \times 10^{41}$\,s$^{-1}$ from the accretion shock on TW~Hya but noted that only $\sim 10^{39}$\,phot\,s$^{-1}$ could reach the disk if the emission is buried under the $\sim3 \times 10^{20}$\,cm$^{-2}$ column of neutral hydrogen gas inferred by X-ray and FUV data. Our upper limit on the $\Phi_{\rm EUV,cm}$ of 1.5$\times 10^{40}$\,s$^{-1}$ suggests that part of the stellar EUV luminosity is indeed absorbed by circumstellar matter before reaching the disk even in this relatively old system. More estimates of the ionizing radiation emitted by pre-main-sequence stars are necessary to evaluate the typical extent of circumstellar extinction. 

\subsection{Evidence for Low-ionization Photoevaporative Winds}\label{sect:low_wind}
All but one (DK~Cha) of the 14 sources studied here have fluxes or upper limits in the \neii{} line at 12.8\,\micron. This transition is relevant to our study because ionized Neon is known to also probe the disk atmosphere of some young stars (e.g. \citealt{sacco12}). In addition, this line is found to be slightly ($\sim$10\,km/s) blueshifted in several disks pointing to unbound gas in a photoevaporative wind (e.g. \citealt{ps09}).  Ne$^+$ could either trace the uppermost layer of the disk surface fully ionized by EUV photons or rather a lower mostly neutral layer where Ne is ionized by 1\,keV X-rays \citep{glassgold07}. The second configuration would imply larger mass loss rates ($> 10^{-9}$\,M$_\odot$/yr) than the first, e.g. the example of TW~Hya in \citet{gorti11} and \citet{pascucci11}. We are now in the position to answer the question: is the $\Phi_{\rm EUV,cm}$ estimated in Sect.~\ref{sect:PhiEUV} large enough, i.e. are there enough ionizing photons reaching the disk atmosphere, to reproduce the observed \neii{} luminosities? 

We first assemble \neii{} fluxes from the literature, giving higher priority to fluxes obtained from spectrally resolved line profiles using ground-based facilities \citep{ps09,pascucci11,baldovin12,sacco12}. When ground-based observations are not available we take the Spitzer/IRS fluxes (\citealt{guedel10,baldovin11,espaillat13}, see also Tables~\ref{tab:sou} and \ref{tab:litsou}). If Neon atoms are ionized by EUV photons, the \neii{} luminosity is directly proportional to the EUV luminosity reaching the disk, hence we can convert the measured \neii{} fluxes into a $\Phi_{\rm EUV,NeII}$. This quantity represents the EUV luminosity necessary to reproduce the observed \neii{} luminosities and can be compared to the upper limits on the EUV luminosity estimated from the excess cm emission ($\Phi_{\rm EUV,cm}$). To compute  $\Phi_{\rm EUV,NeII}$ we use eq.~19 in \citet{hg09} and take the fraction of neon in the singly ionized state equal to unity.

\begin{figure*}
\includegraphics[angle=0,width=\textwidth]{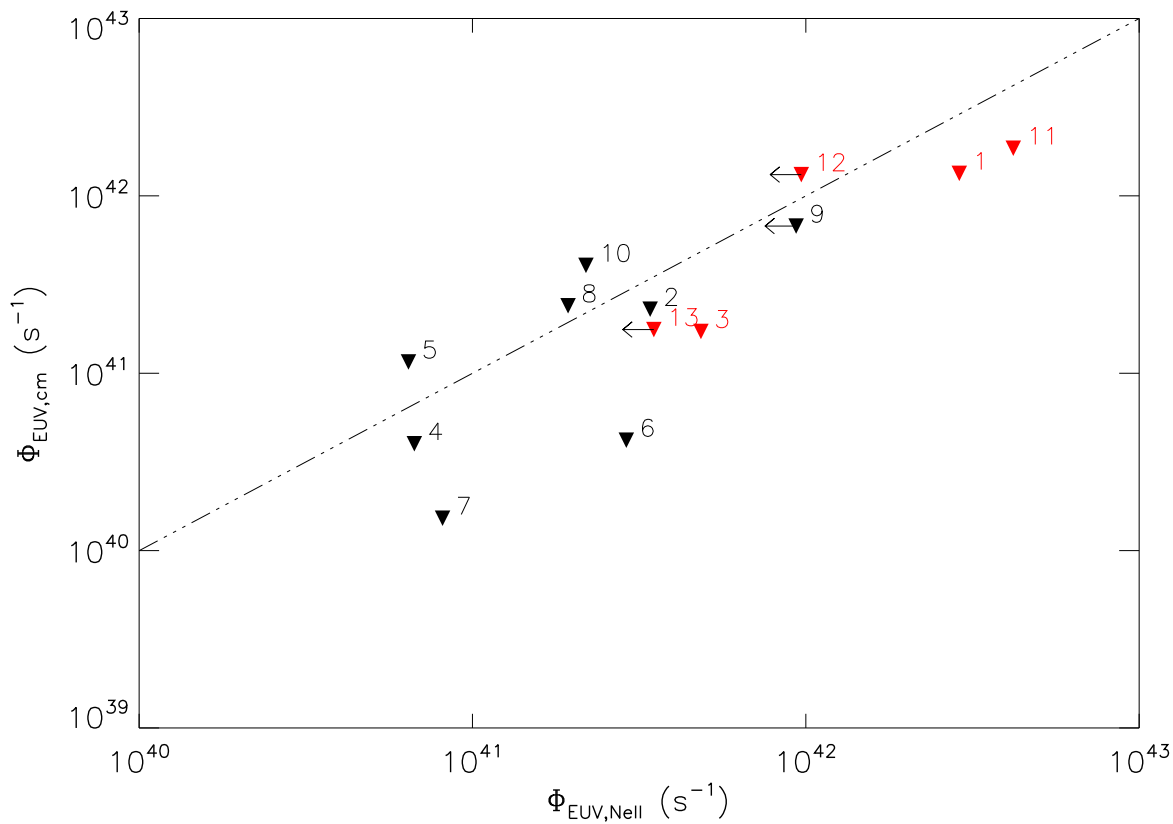}
\caption{Upper limits to the EUV photon luminosity reaching the disk ($\Phi_{\rm EUV,cm}$) as a function of the EUV luminosity needed to reproduce the \neii{} line luminosities or upper limits ($\Phi_{\rm EUV,NeII}$). 
Symbols are as in Fig.~\ref{fig:euv}. The dot-dashed line gives the one-to-one relation.
Several sources lie below this relation suggesting that EUV ionization alone is not sufficient to reproduce the observed \neii{} luminosities.
}
\label{fig:euv_neii}
\end{figure*}

Fig.~\ref{fig:euv_neii} shows $\Phi_{\rm EUV,cm}$ as a function of 
$\Phi_{\rm EUV,NeII}$. The uncertainty in $\Phi_{\rm EUV,NeII}$ is driven by the flux calibration uncertainty at mid-infrared wavelengths, $\sim20$\% for ground-based observations. For three sources (CS~Cha-ID~3, V4046~Sgr-ID~6 and TW~Hya-ID~7) $\Phi_{\rm EUV,cm}$ is clearly not sufficient to reproduce the observed \neii{} luminosities, even when accounting for a factor of 2 uncertainty on the estimated EUV upper limits. Hence, at least in these sources stellar X-rays must contribute to the ionization of Ne atoms. Note that for these three sources there are high-resolution spectra demonstrating that the \neii{} line traces a slow ($\sim 5-10$\,km/s) photoevaporative wind \citep{ps09,pascucci11,sacco12}. This, in combination with the low $\Phi_{\rm EUV,cm}$ upper limits, implies that the \neii{} is not tracing a fully ionized disk layer but rather a lower region only partially ionized by X-ray photons. The presence of a partially ionized and unbound disk layer implies larger mass loss rates than those that can be achieved via EUV-driven photoevaporation alone.

\citet{rigliaco13} recently re-analyzed the low-velocity component of the \oi{} optical forbidden lines from young sun-like stars. Based on the line fluxes, profiles, and peak centroids they argued for the presence of a slow partially molecular photoevaporative flow (driven by X-ray and/or FUV photons) where oxygen is produced by FUV dissociation of OH molecules. Their result also implies higher photoevaporation rates than those produced by EUV-ionization only. The fact that these two independent approaches lead to the same conclusion shows that X-ray and FUV irradiation of the disk surface must be taken into account to estimate realistic photoevaporative mass loss rates.

\section{Conclusions and Implications}\label{sect:conclusions}
This contribution explores the use of cm data to constrain the high-energy radiation reaching the surface of protoplanetary disks and photoionizing H. Because free-free emission from a partially or fully ionized disk layer is optically thin, the free-free flux density is directly proportional to the photon EUV luminosity, $\Phi_{\rm EUV}$. By identifying the cm emission in excess to the thermal dust emission and attributing that to free-free disk emission we obtain upper limits to the $\Phi_{\rm EUV}$ impinging on the disk of 14 young ($\sim2-10$\,Myr) stars. 
Our approach results in two main findings:

\begin{enumerate}
\item  The average $\Phi_{\rm EUV}$ upper limit reaching the disk is $2\times10^{41}$\,s$^{-1}$ in sources without jets and several times lower than $10^{41}$\,s$^{-1}$ for the older systems TW~Hya, MP~Mus, and V4046~Sgr.
\item The inferred $\Phi_{\rm EUV}$ upper limits are not sufficient to reproduce the \neii{} luminosities from three disks, hence stellar X-rays must contribute to the ionization of Ne atoms in these systems.
\end{enumerate}

These two results have interesting implications for our understanding of disk evolution and dispersal. 
The first result shows that the EUV photon luminosity received by the disk is on the low side of the range of stellar EUV luminosities inferred for pre-main sequence stars. 
Such low $\Phi_{\rm EUV}$ luminosities do not appear to be sufficient alone to disperse protoplanetary disks in the timescale that is required by observations. We also note that accounting for gyro-synchrotron and other sources of cm emission would further reduce our estimates (see Sect.~\ref{sect:excess} and Appendix~\ref{appendix:Hubble4}), making it more difficult for EUV photoevaporation alone to clear out protoplanetary material.

The second result demonstrates that, at least in three systems, the \neii{} emission at 12.81\,\micron{} primarily traces a mostly neutral disk region where Ne atoms are ionized by 1\,keV X-ray photons. This, in combination with blueshifts in the peak emission pointing to a wind, indicates that disk gas is photoevaporated deeper in the disk at rates larger than those predicted by EUV irradiation alone. How much higher depends on the relative contribution and evolution of stellar FUV and X-rays in driving and maintaining photoevaporative winds \citep{gorti09,owen11}. In line with the \oi{} 6300\,\AA{} observations \citep{rigliaco13}, our findings demonstrate that star-driven photoevaporation contributes to disk dispersal at higher disk masses than previously thought.

In the next years the synergy between ALMA and sensitive cm interferometers such as the EVLA will enable extending these studies to  larger samples of disks in nearby star-forming regions. These data will show which are the typical upper limits on the EUV luminosity reaching the disk. Such values, in combination with measurements of the mass accretion rate (as suggested in \citealt{owen13}) or direct tracers of the disk ionized surface (as the \neii{} line discussed here) will enable to firmly establish if EUV photons play a minor role in the dispersal of protoplanetary material.

\acknowledgments
The authors thank the anonymous referee for a prompt and useful report.
I.P. thanks Cathie Clarke for stimulating discussion.
I.P., U.G., and D.H. acknowledge support from an NSF Astronomy \& Astrophysics Research Grant (ID: 1312962).

{\it Facilities:} \facility{ATCA}



\appendix

\section{Selected Sources Identified in the ATCA Fields}\label{appendix:ATCAsources}

The 3.3 and 5.5\,cm ATCA images cover a field of over 30 arcminutes, hence several other radio sources are detected in these fields. It is beyond the scope of this paper to characterize all these radio sources but we mention those that are of interest for the interpretation of the excess cm emission and demonstrate how our data reduction recovers known radio sources. The most relevant of these sources is Hubble~4, a young star with no disk but strong cm emission. We dedicate a subsection to this source.

The 3.3 and 5.5\,cm fields of SZ~Cha and CS~Cha cover a well known BL~Lac-type object (PMN~J1057-7724, e.g. \citealt{vv10}). This source dominates the cm emission in its surrounding and even enhances the background at 5.5\,cm at the location of SZ~Cha (see Table~\ref{tab:fluxes}). In the 3.3\,cm and 5.5\,cm image of SR~21 the source located at -100\arcsec{} RA is a known X-ray source in the $\rho$~Ophiuchus cloud core (GDS~J162702.1-241928, \citealt{gagne04}) while the two 3.3\,cm sources at $\sim$\,-50\arcsec{} DEC are close to the young stellar object candidate BKLT~J162707-242009 \citep{bklt01}. Finally, the 3.3 and 5.5\,cm emission at about (+100\arcsec,+25\arcsec) RA, DEC of V4046~Sgr is not associated with a known source. The closest object to this radio emission is a nearby M1-type emission-line star (2MASS~J18142207-3246100) with a large proper motion ($7 \times 40$\,mas/yr, \citealt{z03}).


\subsection{Hubble~4}\label{appendix:Hubble4}
Our ATCA 1.7, 3.3 and 5.5\,cm exposures of V892~Tau also cover the Taurus member Hubble~4 (V1023~Tau, 2MASS~J04184703+2820073). Hubble~4 is a single K7 star classified as a weak-line T~Tauri star based on its low H$\alpha$ EW (-3\,\AA{} which gives an upper limit on the mass accretion rate of $<8\times10^{-9}$\,M$_{\odot}$/yr, \citealt{wg01}). \citet{furlan06} compiled its SED and found no evidence of excess emission out to $\sim$10\,\micron{}. The system is classified as Class~III (no disk) even when extending the SED at 24\,\micron{} with {\it Spitzer}/MIPS \citep{luhman10,rebull10} and millimeter observations place an upper limit on the dust disk mass of only $4\times 10^{-4}$\,M$_\odot$\citep{andrews05}.  Hubble~4 is a strong X-ray emitter (4-6.5$\times 10^{30}$\,erg/s) with modest absorption (3.1$\times 10^{21}$\,cm$^{-3}$, \citealt{guedel07}). The X-ray luminosity is about 8 times larger than the average X-ray luminosity of Taurus sources and, although the X-ray spectrum peaks at $\sim1$\,keV, strong emission can be detected below 0.5\,keV. We use the {\it gaussfit} task in CASA and find that the 1.7, 3.3 and 5.5\,cm emission is not clearly spatially extended beyond the beam (FWHMs of 25\arcsec$ \times 10$\arcsec{}, 36\arcsec$\times 24$\arcsec{} and 63\arcsec$\times 40$\arcsec{} respectively). We measure flux densities of 0.19$\pm$0.02\,mJy, 0.82$\pm$0.08\,mJy and 1.0$\pm$0.1\,mJy at 1.7, 3.3 and 5.5\,cm respectively. Adopting the usual definition of spectral index $\alpha$ as $F_\nu \propto \nu^{\alpha}$, we find $\alpha=$-0.4 using the 3.3 and 5.5\,cm data and  $\alpha=$-1.7 when including the 1.7\,cm datapoint. Such negative spectral indexes exclude thermal free-free emission from a wind or jet and rather point to non-thermal gyro-synchrotron radiation (see discussion in Sect.~\ref{sect:radioslopes}). Recently, \citet{dzib13} have shown that the Gudel-Benz relation between radio and X-ray emission of old magnetically active stars holds even for young stellar objects but with a slightly less steep power law of the form L$_{\rm X}$/L$_{\rm radio}\sim$10$^{14 \pm 1}$. 
The example of Hubble~4 demonstrates that strong cm emission of the order of several hundred $\mu$Jy can be produced by magnetic activity in young stars that lack disks and jets. We note that this emission could account for most of the excess cm emission we measure toward our targets if we scale the flux density of Hubble~4 at 3\,cm by source distance and X-ray luminosity. However, in sources with multiple cm detections the spectral index is mostly positive (see Sect.~\ref{sect:radioslopes}) suggesting that in stars with disks the level of magnetic activity is perhaps lower than in stars without.

\clearpage

\begin{deluxetable}{lccccccccc}
\tabletypesize{\scriptsize}
\rotate
\tablecaption{ATCA sources: properties relevant to this study\label{tab:sou}}
\tablewidth{0pt}
\tablehead{
\colhead{ID} & \colhead{Source} & \colhead{R.A.} & \colhead{Decl.} & \colhead{SpTy} & 
\colhead{Distance} & \colhead{Companion(s)} & \colhead{L$_{\rm X}$} & \colhead{F$_{\rm [NeII]}$} &\colhead{Ref}  \\
\colhead{} & \colhead{} & \colhead{(J2000.0)} & \colhead{(J2000.0)} & \colhead{} & 
\colhead{(pc)} & \colhead{($''$)} & \colhead{(10$^{30}$\,erg/s)} & \colhead{(10$^{-14}$\,erg/s/cm$^{2}$)} &\colhead{}
}
\startdata
1 & V892~Tau & 04 18 40.62 & +28 19 15.5  & B9 & 140 & 0.05, 4 & 9.21-7.94 & 17.9\tablenotemark{w}& 1,2,3,4 \\
2 & SZ~Cha & 10 58 16.77 & -77 17 17.1    & K0 & 160 & 5.3, 12.5 & 0.79 & 1.6 & 5,6,7,8 \\
3 & CS~Cha & 11 02 24.91 & -77 33 35.7    & K6 & 160 & 0.02 & 3.98 & 2.3\tablenotemark{w} & 9,10,7,11 \\
4 & MP~Mus & 13 22 07.53 & -69 38 12.2    & K1IVe & 86 & … & 1.58 & 1.1\tablenotemark{w} & 12,13 \\
5 & SR~21  & 16 27 10.28 & -24 19 12.7    & G3 & 125 & 6.4 & 1 & 0.5\tablenotemark{w} & 14,15,16,13 \\
6 & V4046~Sgr & 18 14 10.48 & -32 47 34.4 & K5 & 73 & 0.04 & 1.2 & 6.6\tablenotemark{w} &17,18,13 \\
\enddata
\tablecomments{J2000.0 coordinates are from the 2MASS All Sky Survey Point Source Catalogue (Skrutskie et al.~2006).
}
\tablerefs{
(1) Briceno et al.~(2002); (2) Kenyon et al.~(2008); (3) Guedel et al. 2007;
(4) Baldovin-Saavedra et al.~(2012); 
(5) Luhman~(2008); (6) Ghez et al.~(1997); (7) Kim et al.~(2009); (8) Espaillat et al.~(2013);
(9) Luhman~(2007); (10) Guenther et al.~(2007); (11) Pascucci \& Sterzik~(2009); 
(12) Mamajek et al.~(2002); (13) Sacco et al.~(2012);
(14) Prato et al.~(2003); (15) de Geus et al.~(1989); (16) Grosso et al.~(2000)
(17) Stempels \& Gahm~(2004); (18) Torres et al.~(2008) }
\tablenotetext{w}{Sources have spectrally resolved \neii{} line profiles with blueshifted peak emission pointing to a slow photoevaporative wind}
\end{deluxetable}

\begin{deluxetable}{llllll}
\tabletypesize{\scriptsize}
\rotate
\tablecaption{Log of the observations\label{tab:log}}
\tablewidth{0pt}
\tablehead{
\colhead{ID} & \colhead{Source} & \colhead{Wavelength} & \colhead{Obs. Date} &
\colhead{Integration\tablenotemark{a}} & \colhead{Gain/Phase}   \\
\colhead{} & \colhead{} & \colhead{(cm)} & \colhead{} & 
\colhead{(min)} & \colhead{Calibrator} 
}
\startdata
1 & V892~Tau & 0.9 & 18Oct2012 & 64 & 0510+180 \\
  &          & 1.7 & 18Oct2012 & 64 & 0333+321   \\
  &          & 3.3, 5.5 &  19Oct2012 & 197 & 0400+258   \\
2 & SZ~Cha & 0.9 & 18-19Oct2012 & 148 &1057-797  \\
  &          & 1.7 & 18Oct2012 & 126 &1057-797  \\
  &          & 3.3, 5.5 & 20-21Oct2012 & 460 &1057-797  \\
3 & CS~Cha & 0.9 & 18Oct2012 & 95 &1057-797  \\
  &          & 1.7 &  19Oct2012 & 63 &1057-797  \\
  &          & 3.3, 5.5  & 19-20Oct2012 & 644 &1057-797   \\
4 & MP~Mus & 0.9 & 18Oct2012 & 63 &J1147-6753  \\
  &          & 1.7 &  19Oct2012 & 32 &J1147-6753  \\
  &          & 3.3, 5.5  & 20-21Oct2012 & 247 &1251-713  \\
5 & SR~21 & 0.9 & 19Oct2012 & 63 &1622-253  \\
  &          & 1.7 &  19Oct2012 & 63 & 1622-253  \\
  &          & 3.3, 5.5  & 19-20Oct2012 & 170 &1622-253   \\
6 & V4046~Sgr & 0.9 & 19Oct2012 & 32 &1759-39  \\
  &          & 1.7 &  19Oct2012 & 32 &1759-39  \\
  &          & 3.3, 5.5  & 21Oct2012 & 126 & 1817-254  \\
\enddata
\tablenotetext{a}{Total time spent on target before flagging and calibration of data}
\end{deluxetable}

\begin{deluxetable}{llcccccccc}
\tabletypesize{\scriptsize}
\rotate
\tablecaption{ATCA continuum fluxes and rms for our science targets. 
We also report the restored beam FWHMs of the cleaned images. \label{tab:fluxes}}
\tablewidth{0pt}
\tablehead{
\colhead{ID} & \colhead{Source} & \colhead{F$_{\rm 0.9\,cm}$ (RMS)} & \colhead{Beam} &
\colhead{F$_{\rm 1.7\,cm}$ (RMS)} & \colhead{Beam} &
\colhead{F$_{\rm 3.3\,cm}$ (RMS)} & \colhead{Beam} &
\colhead{F$_{\rm 5.5\,cm}$ (RMS)} & \colhead{Beam} \\
\colhead{} & \colhead{} & \colhead{mJy (mJy\,beam$^{-1}$)} & \colhead{(arcsec)}&
\colhead{mJy (mJy\,beam$^{-1}$)} & \colhead{(arcsec)}&
\colhead{mJy (mJy\,beam$^{-1}$)} & \colhead{(arcsec)}&
\colhead{mJy (mJy\,beam$^{-1}$)} & \colhead{(arcsec)}
}
\startdata
1 & V892~Tau & 1.8 (0.04) & 10$\times$6 & 1.24 (0.03) & 24$\times$11 & 0.78\tablenotemark{a} (0.02) & 35$\times$24 & 0.57\tablenotemark{b} (0.03) & 60$\times$40  \\
2 & SZ~Cha &  0.17\tablenotemark{c} (0.02) & 8$\times$7 & 0.09\tablenotemark{c} (0.02)& 18$\times$11 & $<$ 0.081 (0.027) & 30$\times$26 &  $<$0.9\tablenotemark{d} (0.3) & 49$\times$42 \\
3 & CS~Cha & 0.35\tablenotemark{c,e} (0.03)  & 8$\times$6 & $<0.12$\tablenotemark{e} (0.04) & 15$\times$11 & 0.38\tablenotemark{b} (0.01) & 30$\times$26 &0.30\tablenotemark{b} (0.04) & 51$\times$43 \\
4 & MP~Mus &  0.80 (0.02) & 8$\times$6 & 0.196 (0.032)& 13$\times$11 & 0.136 (0.013) & 30$\times$24  & $<$0.048 (0.016) & 49$\times$40 \\
5 & SR~21  &  0.13\tablenotemark{c} (0.028) & 6$\times$5 & 0.088\tablenotemark{g} (0.022) & 12$\times$10 & 0.057 (0.011) & 29$\times$19 & $<$0.24 (0.08)& 46$\times$34 \\
6 & V4046~Sgr & 2.4 (0.02) & 6$\times$5 & 0.46 (0.035) & 12$\times$10 & 0.21 (0.011) & 28$\times$19 &$<$0.09 (0.03) & 45$\times$34 \\
\enddata
\tablecomments{All flux densities, except when noted, are computed within the 3$\sigma$ RMS closed contourn. Note that the absolute flux calibration uncertainty (not included in the quoted RMS) is  $\sim$10\% of the measured flux densities.}
\tablenotetext{a}{The target and jet (located $\sim$80$’’$ N) are separated, the reported flux density does not include the jet emission. The 3.3\,cm flux density of the jet is 0.094\,mJy.}
\tablenotetext{b}{The target and jet are not separated, the reported flux density includes the jet emission. Note that in the case of CS~Cha the peak emission is offset from the 2MASS source coordinates suggesting that is dominated by the jet.}
\tablenotetext{c}{Flux densities measured within the 2$\sigma$ RMS contour. The 3$\sigma$ RMS contour flux is less than the one reported by more than the 10\% absolute flux calibration uncertainty suggesting that significant emission is lost with the 3$\sigma$ cutoff (see also text).}
\tablenotetext{d}{High background due to emission from the BL~Lac-type object PMN~J1057-7724 (see Sect.~\ref{appendix:ATCAsources}).}
\tablenotetext{e}{The jet is separated from the star/disk and detected at a $\sim$2$\sigma$ level. The jet flux density within the 2$\sigma$ RMS contour is 0.061\,mJy at 0.9\,cm and 0.096\,mJy at 1.7\,cm. }
\tablenotetext{f}{The jet is not separated from the star/disk system hence the reported flux densities include the jet emission.}
\tablenotetext{g}{We discarded the 19\,GHz dataset because it has a rms that is 50\% higher than the 17\,GHz dataset.}
\end{deluxetable}

\begin{deluxetable}{lccccccc}
\tabletypesize{\scriptsize}
\tablecaption{Literature sources: properties relevant to this study\label{tab:litsou}}
\tablewidth{0pt}
\tablehead{
\colhead{ID} & \colhead{Source}  & \colhead{SpTy} & \colhead{Distance}  
& \colhead{L$_{\rm X}$} & \colhead{F$_{\rm [NeII]}$} & \colhead{Jet?}  &
\colhead{Ref}  \\
\colhead{} & \colhead{} & \colhead{} & \colhead{(pc)} & 
\colhead{(10$^{30}$\,erg/s)} & \colhead{(10$^{-14}$\,erg/s/cm$^{2}$)} &\colhead{}
&\colhead{}
}
\startdata
7 & TW~Hya   & K6  & 51 & 1.5 & 3.8\tablenotemark{*,w} & n & 1,2,3,4 \\
8 & GM~Aur   & K7  & 140 & 1.6 & 1.2& n&5,6,7 \\
9 & RY~Tau   & G1  & 140 & 5.5 & $<$5.8 & n&5,6,8,9  \\
10 & T~Cha   & K0 & 110 & 43. & 2.2\tablenotemark{w}& n & 10,11,12 \\
11 & DG~Tau  & K6 & 140 & 0.55 & $26$ & yes & 5,7,6 \\
12 & HL~Tau  & K7  &  140 & 3.8 & $<6$ & yes & 13,6,8,14\\
13 & UZ~TauE & M1+M4 & 140 & 0.9 & $<$2.2 & yes& 15,6,8,14 \\
14 & DK~Cha  & F0  & 178 & 13& — & yes & 16,17,18\\
\enddata
\tablerefs{
(1) Torres et al.~(2006); (2) Mamajek~(2005); (3) Brickhouse et al.~(2010); (4) Pascucci et al.~(2011);
(5) Furlan et al.~(2011); (6) Kenyon et al.~(2008); (7) Guedel et al.~(2010);
(8) Guedel et al.~(2007); (9) Baldovin-Saavedra et al.~(2012);
(10) Torres et al.~(2008); (11) Sacco et al.~(2014); (12) Pascucci \& Sterzik~(2009);
(13) Luhman et al.~(2010); (14) Baldovin-Saavedra et al.~(2011); (15) Prato et al.~(2002);
(16) Spezzi et al.~(2008); (17) Hamaguchi et al.~(2005); (18) Ubach et al.~(2012)}
\tablenotetext{*}{Average of several values}
\tablenotetext{w}{Sources have spectrally resolved \neii{} line profiles with blueshifted peak emission pointing to a slow photoevaporative wind}

\end{deluxetable}

\begin{deluxetable}{lllll}
\tabletypesize{\scriptsize}
\rotate
\tablecaption{Millimeter and centimeter measurements compiled from the literature.\label{tab:litval}}
\tablewidth{0pt}
\tablehead{
\colhead{ID} & \colhead{Source} & \colhead{$\lambda$} & 
\colhead{Flux} & \colhead{Ref}\\
\colhead{} & \colhead{} & \colhead{(cm)} & 
\colhead{(mJy)} & \colhead{}
}
\startdata
1 & V892~Tau & 0.085, 0.13, 0.29  & 638, 234, 55,  & 1, 2 \\
2 & SZ~Cha & 0.087, 0.12, 0.32, 0.68 & 314, 77.5, 5.8, 0.7  & 3, 4\\
3 & CS~Cha & 0.087, 0.12, 0.32, 0.67, 0.7, 1.56, 1.56, 1.76, 1.76 & 197, 128, 9.4, 1.4\tablenotemark{*}, 1.3\tablenotemark{*}, 0.4, $<$0.2, 0.5, $<$0.3  & 3, 4   \\
4 & MP~Mus & 0.12, 0.3, 1.2, 3.0, 6.0 & 207, 22, 0.41, $<$0.2, $<$0.25  & 5, 6\\
5 & SR~21  & 0.088, 0.1, 0.13, 0.33,   & 400, 220, 150, 4.2,  & 7, 2, 8, 9  \\
6 & V4046~Sgr & 0.08, 0.11, 0.13 & 770, 415, 360,  & 10, 11, 12 \\
\hline
7 & TW~Hya & 0.087, 0.11, 0.13, 0.34, 0.7, 3.5, 4.1, 6.3 & 1340, 874, 570, 41, 8, 
0.26, 0.145, 0.075 &  13\\
8 & GM~Aur & 0.1, 0.11, 0.13, 0.2, 0.27, 0.7, 1.7 & 423, 380, 253, 37, 28, 1.1, 0.13  & 2, 14, 15 \\
9 & RY~Tau & 0.089, 0.11, 0.12, 0.13, 0.2, 0.7, 1.3, 2.0, 3.6  & 499, 280, 212, 193, 52, 2.97, 0.92, 0.63, 0.31 & 16, 14 \\
10 & T~Cha & 0.12, 0.32, 0.68, 1.5, 3.0, 6.0 & 105, 6.4, 0.7, 0.3, $<$0.3, 0.3 & 4  \\
11 & DG~Tau & 0.11, 0.12, 0.13, 0.2, 0.27, 0.3, 1.3, 2.0, 3.6 & 570, 420, 443, 140, 59, 57, 2.17, 1.33, 1.27 & 14  \\
12 & HL~Tau & 0.11, 0.12, 0.13, 0.2, 0.27, 0.3, 1.3, 2.0, 3.6 & 1110, 961, 879, 
161, 107, 74, 1.63, 0.88, 0.57 &  14 \\
13 & UZ~TauE & 0.1, 0.12, 0.13, 0.27, 0.31, 1.3, 2.0, 3.6 & 333, 170, 137, 26, 
14, 0.77, 0.48, 0.1 & 14 \\
14 & DK~Cha & 0.12, 0.32, 0.68, 1.5, 3.0, 6.0 & 680, 50, 6.6, 1.6, 0.8, $<$0.3  &  4 \\
\enddata
\tablerefs{
(1) Andrews et al.~(2005); (2) Ricci et al.~(2012);
(3) Belloche et al.~(2011); (4) Ubach et al.~(2012);
(5) Carpenter et al.~(2005); (6) Cortes et al.~(2009);
(7) Andrews et al.~(2011); (8) Andre \& Montmerle~(1994); (9) Ricci et al.~(2010a);
(10) Jensen et al.~(1996); (11) Oberg et al.~(2011); (12) Rodriguez et al.~(2010);
(13) Menou et al.~(2014); (14) Rodmann et al.~(2006); (15) Owen et al.~(2013);
(16) Andrews et al.~(2013)  }
\tablenotetext{*}{Mean of several measurements}
\end{deluxetable}

\begin{deluxetable}{llcc}
\tabletypesize{\scriptsize}
\tablecaption{Upper limits on the EUV photon luminosity ($\Phi_{\rm EUV,cm}$) reaching the disk. These upper limits could be at most a factor of 2 higher.\label{tab:euv_inferred}}
\tablewidth{0pt}
\tablehead{
\colhead{ID} & \colhead{Source} & \colhead{$\Phi_{\rm EUV,cm}$} & 
\colhead{$\lambda_{\rm EUV,cm}$} \\
\colhead{} & \colhead{} & \colhead{(10$^{41}$\,s$^{-1}$)} & 
\colhead{(cm)}
}
\startdata
1 & V892~Tau & 13.4& 5.5\\
2 & SZ~Cha & 2.3 & 1.7\\
3 & CS~Cha &  1.7 & 1.7\\
4 & MP~Mus &  0.4 & 5.5\\
5 & SR~21  &  1.2  & 3.3\\
6 & V4046~Sgr & 0.4 & 5.5\\
\hline
7 & TW~Hya & 0.15 & 6.3\\
8 & GM~Aur & 2.4 & 1.7\\
9 & RY~Tau & 6.8  & 3.6\\
10 & T~Cha & 4.1 & 1.5\\
11 & DG~Tau & 18.5 & 1.3\\
12 & HL~Tau & 13.2 & 3.6\\
13 & UZ~TauE & 1.8 & 3.6\\
14 & DK~Cha & 11.1 & 6.0\\
\enddata
\end{deluxetable}


\end{document}